\documentclass[aps,prd,preprint,groupedaddress]{revtex4}
\usepackage[dvipdfmx]{graphicx}
\usepackage{ulem}

\begin{document}


\title{A Hydrodynamical Study on the Conversion of Hadronic Matter to Quark Matter: I. Shock-Induced  Conversion}

\author{Shun Furusawa}
\email{furusawa@cfca.jp}
\affiliation{Center for Computational Astrophysics, National Astronomical Observatory of Japan, Osawa, Mitaka, Tokyo, 181-8588, Japan}
\affiliation{Advanced Research Institute for Science and Engineering, Waseda University, 3-4-1 Okubo, Shinjuku, Tokyo 169-8555, Japan}
\author{Takahiro Sanada}
\affiliation{Department of Science and Engineering, Waseda University, 3-4-1 Okubo, Shinjuku, Tokyo 169-8555, Japan}
\author{Shoichi Yamada}
\affiliation{Advanced Research Institute for Science and Engineering, Waseda University, 3-4-1 Okubo, Shinjuku, Tokyo 169-8555, Japan}


\date{\today}

\begin{abstract}
We study transitions of hadronic matter (HM) to 3-flavor quark matter (3QM) locally, regarding the conversion processes as combustion and describing them hydrodynamically. 
Not only the jump condition on both sides of the conversion front but the structures inside the front are also considered by taking into account what happens during the conversion processes on the time scale of weak interactions as well as equations of state (EOS's) in the mixed phase. 
Under the assumption that HM is metastable with their free energies being larger than those of 3QM but smaller than those of 2-flavor quark matter (2QM),
we consider the transition via 2QM triggered by a rapid density rise in a shock wave. 
Based on the results, we discuss which combustion modes (strong/weak detonation) may be realized. 
HM is described by an EOS based on the relativistic mean field theory and 2, 3QM's are approximated by the MIT bag model. 
We demonstrate for a wide range of bag constant and  strong coupling constant  in this combination of EOS's that   the  combustion may occur
 in the so-called endothermic regime, in which the Hugoniot curve for combustion runs below the one for the shock wave in $p-V$ plane, and which has no terrestrial counter part.
Elucidating the essential features in this scenario
first by a toy model, we then analyze more realistic models. 
We find that 
strong detonation always occurs. 
Depending on the EOS of quark matter (QM) as well as the density of HM and the Mach number of the detonation front, deconfinement from HM to 2QM is either completed or not completed in the shock wave. In the latter case, which is more likely if the EOS of QM ensures that deconfinement occurs above the nuclear saturation density and that the maximum mass of cold quark stars is larger than $2M_{\odot}$, the conversion continues further via the mixing state of HM and 3QM on the time scale of weak interactions.

\end{abstract}


\maketitle

\section{Introduction \label{intro}}
The hadronic equation of state (EOS) at supra-nuclear densities ($\gtrsim 2.8 \times 10^{14}$~g/cm$^{3}$), which are believed to prevail at the central region of neutron stars, is still highly uncertain. 
In fact, not only nucleons but hyperons may also exist or Bose-Einstein condensations of mesons like pions or kaons might take place (see, for example, \cite{baym2006}). It is also possible that quark matter (QM) prevails over a substantial part of neutron star (such a star is referred to as a hybrid star) and, indeed, the entire star may consist of deconfined quarks \cite{itoh70,alcock} if 3QM, which is referred to as strange quark matter (SQM)
in this case, is the most stable state at zero pressure. 

SQM is a bulk QM, which is composed of roughly the same numbers of up, down and strange quarks (plus a small fraction of electrons for charge neutrality) and is hypothesized to be the true ground state of strong interactions \cite{witten}.
SQM can have  various baryon numbers from $A \sim 100$ to $A \sim 10^{57}$. 
Macroscopic quark nuggets are generically called nuclearites and those with a small baryon number ($A < 10^{6}$), in particular, are referred to as strangelets. Such quark nuggets may have been produced in the early universe. 
It may be also possible that they are produced in the center of neutron stars still in the present universe.
If they are released into galaxies by, for example, collisions of compact stars \cite{madsen,friedman} or supernova explosions \cite{vucetich}, elastic or quasi elastic collisions of nuggets with atoms and molecules in the traversed matter might be detected 
experimentally \cite{rujula} although such attempts have been so far unsuccessful yet (see, for example, \cite{cecchini}).

If SQM is formed in a neutron star by some mechanism, which is referred to as seed in the following, HM will be subsequently converted to SQM at the boundary of HM and SQM and the entire star (possibly except for a thin crust at the surface, since the Coulomb barrier may prevent the conversion below the neutron-drip density \cite{alcock}) will be eventually composed of SQM and is called the strange star. The conversion will liberate a large amount of gravitational energy $\sim 10^{53}$erg, which is comparable to that of a neutron star and has been frequently advocated as a possible energy source of gamma ray bursts \cite{bombaci04,mallick14}. 
Possible effects of the conversion on supernova explosions \cite{benvenuto89,benvenuto,lugones94,gentile,dai,sagert2009,sagert2010,fischer2010,fischer2011,fischer2011b} as well as on the subsequent cooling of neutron stars have been also investigated by many authors \cite{dexheimer2011,kang2010}.

Various mechanisms of the seed formation in the neutron star were discussed in \cite{alcock}. The scenarios can be divided into two categories depending on whether the seed SQM is produced inside the neutron star itself or not. 
Included in the first category are conversions via 2QM, clustering of lambda hyperons, neutrino sparking and so on. 
In particular, if the density in a neutron star reaches a certain critical value, $\rho _c$, by, e.g., a spin down of neutron star \cite{glendenning,yasutake05}, 2QM becomes energetically favored and the deconfinement of up and down quarks will occur via strong interactions alone. 
Then SQM, which is by definition more stable than 2QM, will be produced via weak interactions,  the process that can  be regarded as $\beta$-equilibration. In the second category, on the other hand, it is supposed that SQM is produced somewhere else and is later trapped by neutron stars. 
In one of such scenarios the seed was formed in the early universe and propagates through space. 
The formation rate of strange stars was estimated and it was argued that all neutron stars are in fact strange stars \cite{olinto1991}. 
Another possibility is that strange stars, which have been already formed, may somehow eject a fraction of SQM into space, which might later become a seed SQM in another neutron star. 

If SQM is the true ground state of strong interactions, HM should be metastable and their decay is avoided by the fact that intermediate states with smaller fractions of strangeness are unstable compared with HM.
The conversion of the metastable state to the truly stable state separated by unstable states can be regarded as combustion: HM is a fuel and SQM is an ash; 
there is a conversion front in between, in which the mixtures of fuel and ash exist and the conversion process takes place. This conversion region is very thin compared with macroscopic scales, e.g., stellar radii.
In the hydrodynamical description of terrestrial combustions \cite{williams,landau,barton1973}, the fuel and ash are related with each other by the so-called Hugoniot relation and there are in general four combustion modes, strong/weak detonation/deflagration, of which the strong deflagration is thought to be unrealizable. 
Which mode actually occurs is determined by the conversion mechanism and parameters involved.
 In this paper and its sequels, we model the structure of conversion front, adopting the hydrodynamical description. 
We then discuss which combustion mode is likely to be realized. 

The propagation speed of the conversion front and the time scales of the conversion of entire neutron stars have been estimated by many researchers. 
Olinto \cite{olinto} was the first to infer the front velocity under the assumption that the conversion is induced by diffusions of seed SQM. Ignoring the equations of motion and assuming a constant velocity of SQM, she obtained the front velocity of $10^0 -10^4$km/s depending on the  critical fraction of strange quarks, at which 3QM becomes more stable than HM. 
Heiselberg derived the transport equations for up, down and strange quarks, solved them analytically, assuming a local thermal equilibrium and a constant total pressure as well as marginal flammability, and found an even smaller front speed of $\sim 10$m/s \cite{heiselberg}.
These results correspond essentially to the (very slow) weak deflagration 
among the four combustion modes
mentioned above. Employing the estimation of the front velocity similar to the one given by Olinto~\cite{olinto} and solving an equation for hydrostatic configurations of hybrid stars, Olesen et al. \cite{olesen} estimated the conversion time scale for a whole neutron star. 
They adopted Bethe-Johnson's equation of state \cite{bethe1974} for neutron matter and the MIT bag model for QM. In solving the hydrostatic equations they assumed pressure and chemical equilibrium of up and down quarks and local charge neutrality at the conversion front. For various bag constants and initial temperatures, they found the time scale of complete conversion ranges from 0.1 seconds to a few minutes. 
Benvenuto at al. \cite{benvenuto89,benvenuto} were the first to investigate the fast combustion mode known as detonation, which is actually the focus  in  this paper. 
They solved the relativistic  Hugoniot relations with the Bethe-Johnson's EOS for neutron matter and the MIT bag model for QM and discussed  the possible implications for supernova explosions.

Niebergal et al.~\cite{brian} investigated the effects of global dynamics on the front velocity, numerically solving hydrodynamical equations in spherical symmetry together with neutrino emissions from $\beta-$equilibrating reactions as well as diffusions of strangeness. 
Although the equation of state for HM was approximated by the bag model for 2QM, they found that the global dynamics has a non-negligible effect and obtained the front velocity of 0.002-0.04 times the speed of light, much faster than the previous estimates neglecting the dynamics. 
Effects of turbulence on the deflagration front velocity was studied numerically in 3D large eddy simulations by \cite{herzog,pagliara13}. 
The time scale to convert the whole star is  the order of a few milliseconds in their simulations.
They claimed that the front velocity is enhanced substantially compared with the laminar case. 
Advocating the picture that the conversion of neutron stars proceeds via two steps, in which a neutron star is first converted to a quark star composed of 2QM, which is then converted to a strange star in the second step, Bhattacharyya et al.~\cite{bhattacharyya} estimated the time scale of each step. They found that the first conversion took about a millisecond whereas the second step proceeds over a hundred seconds. 
Very recently Mishustin et al.  \cite{mishustin14} pointed out  that the detonation front velocity may  increase or decrease depending on the velocity of incoming HM.

Some authors have paid their attention to the hydrodynamical properties of the conversion such as the combustion modes. 
Considering the jump condition at the combustion front and the flammability condition, for example, Cho et al.~\cite{cho} claimed that neither strong detonation nor weak deflagration is possible whereas weak detonation is possible under certain circumstances and that the burning is most likely to be unstable. 
The conclusion was challenged by Tokareva et al.~\cite{tokareva}, who insisted that all the combustion modes are possible. Lugones et al.~\cite{lugones94} also studied the flammability condition as well as the effect of hydrodynamical boundary condition on the front velocity and suggested that the actual combustion mode should be strong detonation. 
On the other hand, employing a more realistic EOS for HM and taking into account the mixed phase by the so-called Gibbs condition, Drago et al.~\cite{drago} concluded that the conversion process always occurs via deflagration even if one considers possible enhancement of the front velocity by hydrodynamical instabilities. 
They contended further that the combustion occurs in the strong-deflagration regime. 
The short list of the preceding studies given here is clear evidence that no consensus has been reached yet on which combustion modes are realized and further investigations are needed \cite{horvath2010}. 
That is the aim of this and following papers.

In this series of papers, we study the transitions of HM to 3QM locally from the hydrodynamical point of view.
We assume that HM is metastable and has free energies that are higher (or less stable) than those of 3QM but are lower (or more stable) than those of 2QM. Note that it is not necessarily assumed that 3QM is absolutely stable, i.e., the most stable at zero pressure, although the SQM hypothesis is 
included as a subset. 
The main difference from the previous studies is that not only the Hugoniot relation between HM on one side of the conversion front and 3QM on the other side but the structures inside the front are also considered by taking into account what will happen during the conversion processes as well as equations of state (EOS's) in the mixed phase. 
The length scale of our interest is the one determined by weak interactions, which is actually the width of the conversion front and much larger than the mean free path for strong interactions whereas it is much smaller than the macroscopic scales, e.g., stellar radii. 
This justifies the employment of the hydrodynamical description in plane symmetry. 
We discuss which combustion modes (strong/weak detonation/deflagration) are likely to be realized for the following two scenarios: (1) the transition via 2QM triggered by a rapid increase in density owing to the passage of a shock wave and (2) the conversion induced by diffusions of a seed 3QM.  
We focus on the former case in this paper and the latter will be reported in the next paper.

It should be noted that our analysis is local, i.e., only the region that just covers the conversion front is taken into account. This is in sharp contrast to the global study of the conversion of entire neutron stars by simulations  \cite{herzog,pagliara13}. The two methods are complementary to each other in fact. In the former one can consider in detail, albeit phenomenologically, what is happening inside the conversion region, which cannot be resolved by global simulations. On the other hand, possible back reactions and global configurations as well as boundary conditions cannot be taken into account in the local analysis. As a matter of fact, although the front velocity is a free parameter in our analysis, it is actually determined uniquely by the global configuration and boundary conditions in the conversion of neutron stars to quark stars. We hence believe that the correct picture of the conversion of neutron stars to quark stars is obtained only with a proper understanding of both of these aspects. In this paper we try to list up all possible structures but make no attempt to claim which ones are more likely than others to be realized in the actual conversion. In this sense, the conditions we consider in this paper are just necessary conditions but not sufficient ones.

We also stress in this series of papers that for the combination of realistic baryonic EOS's
 such as the one we employ in this paper and the bag model EOS's for QM with a wide range of bag and strong coupling constants, 
the combustions may occur in the so-called endothermic regime,
 in which the Hugoniot curve for combustion runs below the one for shock wave. 
Such a combustion has no terrestrial counterpart \cite{williams, barton1973} and has been discarded in the previous papers exactly because it is endothermic \cite{cho, lugones94}. 
We emphasize, however, that there is no reason in fact to throw it away. 
 In some papers \cite{herzog,drago15}, it is  argued that
 combustion can occur only in the exothermic regime because of the so-called Coll's condition \cite{coll76,anile89}:
 the internal energy  of fuel, i.e., HM in the present case, should be larger than that of 
ash, namely 3QM in  $\beta$-equilibrium in our case, for the density and pressure of the initial state.
Note, however, that the Coll's condition is neither a necessary condition nor a sufficient one but a hypothesis or assumption actually, to which no physical justification is given. In fact it is almost equivalent to the requirement that the Hugoniot curve for combustion should run above that for shock wave. The Coll's condition is hence nothing but the conclusion itself. It should be emphasized that it is not the internal energy but the free energy that determines thermodynamic evolutions. As long as there is no obstacle in between the initial and final states such as an intermediate state with a higher free energy, reactions proceed spontaneously to realize the free-energy minimum. This is exactly the case in the shock-induced conversion considered in this paper: the shock wave compresses HM to the density, at which the intermediate 2QM has the same free energy as HM and is no longer an obstacle for conversion; further shock compression hence ensures deconfinement from HM to 2QM; 3QM, by definition, has a lower free energy than 2QM because the former is the destination of  $\beta$-equilibration. Note that there is nothing to stop this process even if the internal energy of 3QM is higher than that of HM at the same density and pressure.
Note also that the terminology of "exothermic/endothermic combustion" is somewhat misleading, since it does not necessarily correspond to heat production/absorption. 
As shown later neither density nor pressure is constant in the conversion. 
In the following we hence consider both the exothermic and endothermic regimes on the same basis and discuss possible implications thereof.

The outline of the paper is as follows. In Sec.~\ref{scenario}, we describe the scenario more in detail. 
Presenting the Hugoniot curves for the combination of EOS's of our choice in this paper, we demonstrate that the combustion occurs in the endothermic regime more often than not. 
To expedite the understanding of the main results, we present in Sec.~\ref{toy} some fundamental features of the combustion fronts both in the exothermic and endothermic regimes for a toy model that captures the essential ingredients of the more realistic models described in Sec.~\ref{model}. The basic equations and EOS's used for QM, HM and the mixed phase in the combustion front
are given in Sections~\ref{modeleos}, \ref{modelstate}, \ref{model_crid} and \ref{modeleq}, respectively and the main results are presented in Sec.~\ref{result}. 
The paper is concluded with the summary and discussions in Sec.~\ref{conclusion}.

\section{Scenarios \label{scenario}}
\subsection{free energies and schematic pictures of conversions\label{cartoon_energy}}

\begin{figure}[t]
\includegraphics[width=16cm]{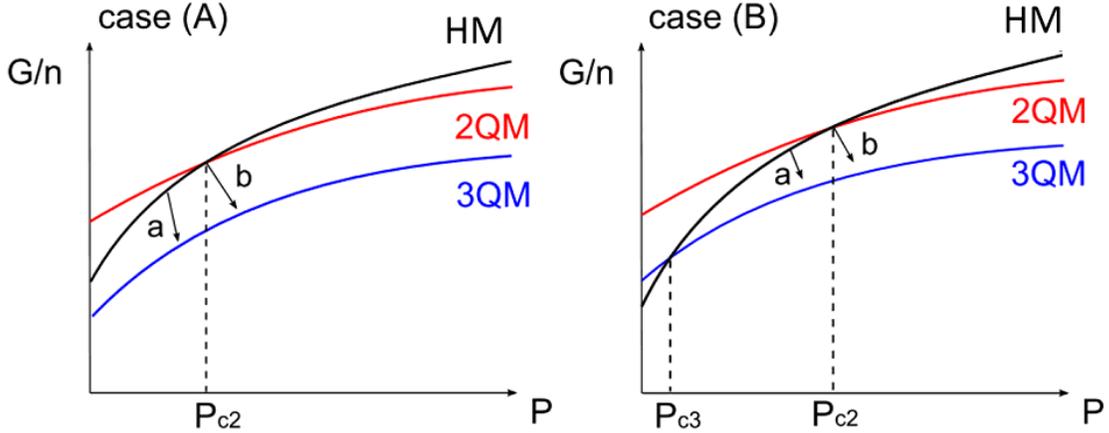}
\caption{Schematic pictures of the Gibbs free energy per baryon. The left panel corresponds to the strange quark matter hypothesis. 
The points, at which the free energy of HM coincides with those of 3QM and 2QM, are denoted by $P_{c3}$ and $P_{c2}$, respectively. 
\label{fig_scenario_energytoy}}
\end{figure}

The situations we have in mind in this paper are best illustrated in Fig.~\ref{fig_scenario_energytoy}, in which the Gibbs free energies per baryon are schematically displayed for HM and 2, 3QM's at zero temperature as a function of pressure~\cite{horvath88}. 
The matter having the lowest free energy per baryon for a given pressure is the most stable there. The left panel corresponds to the SQM hypothesis, under which 3QM is the most stable down to zero pressure. 
In the right panel, on the other hand, HM is the most stable at zero pressure and at a certain pressure shown as $P_{c3}$ in the figure, 3QM takes its place, having the lowest free energy per baryon. 
At still higher pressures ($P \ge P_{c2}$ in the figure), even 2QM becomes more stable than HM in both cases. 
It is important to recognize that the two cases are essentially identical in the vicinity of $P_{c2}$. If the conversion from metastable HM to truly stable 3QM occurs between $P_{c3}$ and $P_{c2}$ as shown by arrows $a$, the two cases do not differ qualitatively. In the SQM hypothesis (case (A)), the diffusion-induced conversion can take place anywhere below $P_{c2}$ 
whereas it is forbidden below $P_{c3}$ in case (B). The shock-induced conversion, on the other hand, occurs at $P_{c2}$ as
indicated by arrows $b$ in both cases.

As already mentioned, we consider the shock-induced 
conversion alone in this paper 
and the diffusion-induced conversion will be discussed in the sequels.
The transition from HM to 3QM is supposed to occur via 2QM in this scenario: the density and 
temperature rise in the shock front rapidly on the time scale of  strong interactions and reach the point, where the free energies of HM and 
2QM coincide with each other and the deconfinement of nucleons to up and down quarks takes place as a phase transition in equilibrium. 
Once 2QM is formed, there is no obstacle other than the finite mass of strange quark to prevent the  $\beta$-equilibration to 3QM from 
proceeding spontaneously. This is in sharp contrast  to the direct transition from HM to 3QM, in which small fractions of strangeness are 
energetically disfavored. Note that the chemical potentials of up and down quarks are close to the strange quark mass in the
typical situations and the time scale of  $\beta$-equilibration will be $\sim 10^{-8}$s~\cite{cho}, which is much shorter than the hydrodynamical 
time scale ($\sim 10^{-3}$s) in neutron stars. This implies that the deconfinement and subsequent  $\beta$-equilibration are lagged by 
$\sim 10^{-8}$s at most and the width of the conversion region is not larger than $\sim 10^2$cm, which is much greater than the typical 
length  scale of strong interactions $\sim $fm, but still much smaller than the typical macroscopic scale such as a neutron star radius 
$\sim 10$km. The conversion region can be hence regarded as a single entity. 

The shock wave may be produced when spin-down of or matter-accretion onto a neutron star increases the central density and temperature 
to the values, at which the free energy per baryon of HM is equal to that of 2QM and the deconfinement of up and down quarks occurs 
spontaneously. Then the $\beta$-equilibration to 3QM ensues immediately. 
Once the shock wave is generated, the conversion process will be self-sustained as long as the shock compression gives densities and temperatures high enough for 
deconfinement. It is then obvious that the conversion is initiated at $P_{c2}$ in Fig.~\ref{fig_scenario_energytoy} for both cases (A) and (B). The 
conversion is schematically expressed by arrows $b$ in the figure. 

\begin{figure}[t]
\includegraphics[width=7.5cm]{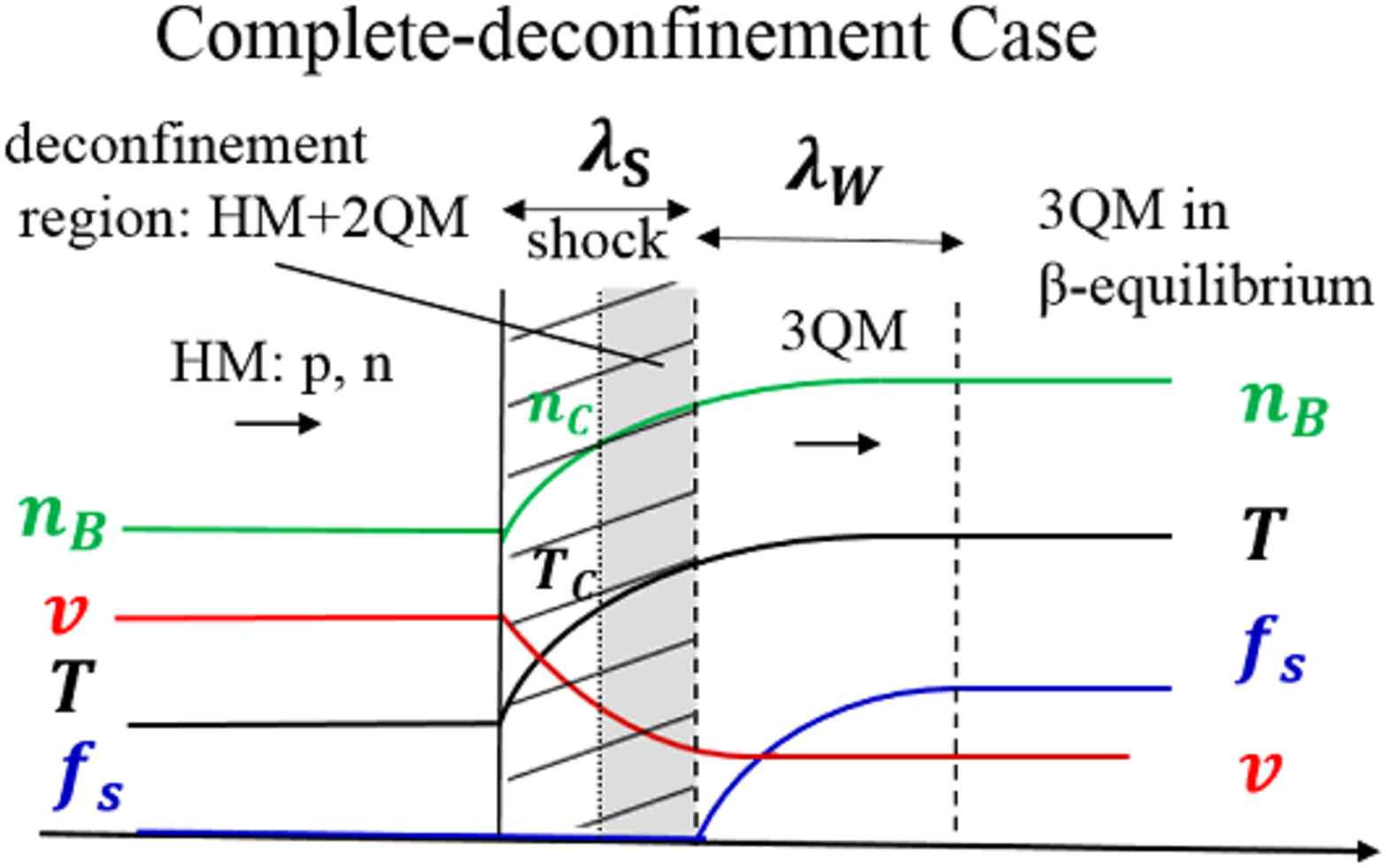}
\includegraphics[width=7.5cm]{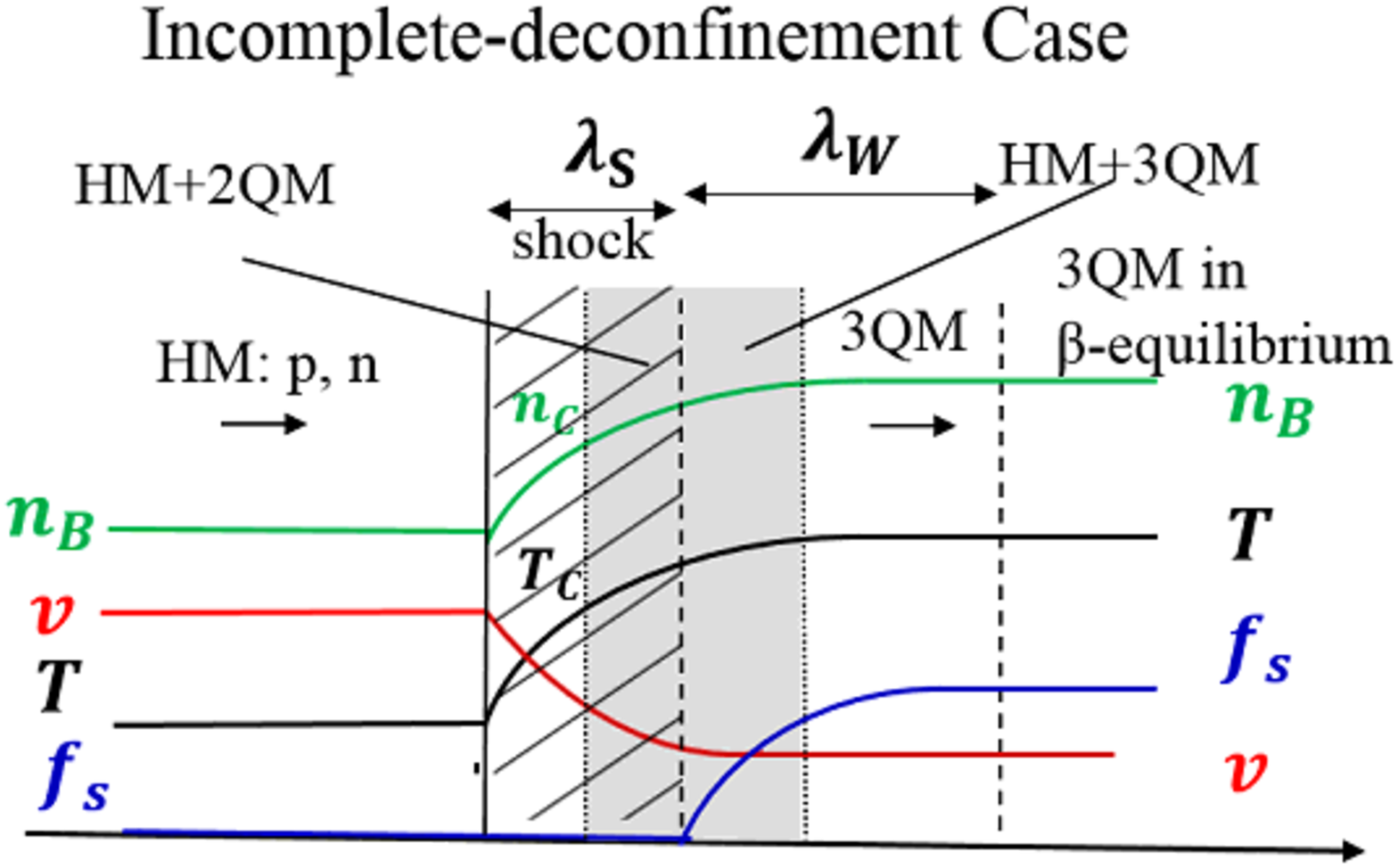}
\caption{The schematic pictures of the shock-induced conversion regions for the complete-deconfinement case (left panel) and the incomplete-deconfinement case (right panel). 
HM composed of protons and neutrons, which are denoted as : $p$, $n$, respectively, in the drawing, is put on 
the left end and 3QM made up of up, down and strange quarks 
 occupies 
the opposite end. They are flowing rightwards in this front-rest frame. The shaded regions stand for the deconfinement regions,
the widths of which are exaggerated in this picture.  The hatches 
correspond to the shock wave. Note that the deconfinement commences inside the shock wave but may end either inside the shock wave (left panel) or outside (right panel).
The lines labeled as $v$, $n_{B}$, $T$ and $f_s$ represent the velocity, baryon number density, 
temperature and fraction of strange quarks, respectively. Leptons are not shown in this picture. 
See the text for the meanings of $\lambda_s$, $\lambda_w$,  $n_c$ and $T_c$.}
\label{fig_structure}
\end{figure}

The structure of the conversion region for this scenario is also composed of a couple of parts and schematically depicted in Fig.~\ref{fig_structure}. The conversion is initiated by the hydrodynamical shock wave. The density and temperature 
increase in HM by shock compression. When they reach the values (denoted as $n_c$ and $T_c$ in the figure), at which the free energy  per 
baryon of HM is equal to that of 2QM, the deconfinement of up and down quarks takes place spontaneously. It should be noted that the 
shock compression occurs on the time scale of strong interactions, $t_s$, and the shock width is $\lambda_s \lesssim c t_s$, in which c is the light speed.
Since $t_s$ is much shorter than the time scale of weak interactions, $t_w$, there is essentially no weak interactions
up to this point. 
As a result, the ratio of the number density of up quarks to that of down quarks is unchanged and no strange quark is present in the shock 
wave. Then the ordinary first order phase transition in equilibrium converts HM to 2QM through a mixed phase. This transition may or may 
not be completed in the shock wave. If it is indeed finished and the volume fraction of 2QM becomes unity by the end of the shock wave, 
the $\beta$-equilibration to 3QM occurs everywhere simultaneously. If the mixed state of
HM and 2QM remains at the end of shock wave, on the other hand, the conversion of 2QM to 3QM via $\beta$-equilibration commences
only in the volume occupied by QM. As this irreversible conversion proceeds in QM, the volume fraction of HM becomes smaller and eventually 
QM prevails over the whole volume and approaches the asymptotic state of 3QM in complete $\beta$-equilibrium. In both cases,
the $\beta$-equilibration proceeds on the time scale of weak interactions, $t_w$, and the conversion region extends over the length
of $\lambda_w \sim c t_w$, which is much wider than $\lambda_s$, and the shock wave is magnified unproportionately in the figure.

%

\subsection{Hugoniot curves \label{jump}}
Although the main goal of this paper is to discuss the structures in the conversion region, we consider in this subsection the relation 
connecting the initial state of HM with the final state of 3QM. 
HM just prior to the conversion is metastable and 3QM in  $\beta$-equilibrium is the truly stable state. Then the transition is quite similar to combustions, in which 
fuels are metastable states whose spontaneous conversion to more stable ashes is prevented by the existence of intermediate states with positive activation energies. 
The relation between the 
initial and final states is obtained from the conservations of baryon number, momentum and energy, which are expressed in the front-rest frame as  
\begin{eqnarray}
\rho_i v_i &=& \rho _f v_f, \label{eq:rankine_m}\\
P_i+\rho v_i^2 &=& P_f +\rho _f v_f ^2, \label{eq:rankine_p}\\
h_i+ \frac{1}{2} v_i^2 &=& h_f + \frac{1}{2}v_f ^2, \label{eq:rankine_e}
\end{eqnarray}
where the subscripts $i$ and $f$ represent the initial and final states, respectively, and $\rho$, $v$, $P$ and $h$ stand for the baryon mass density, velocity, pressure and specific enthalpy, respectively. From Eqs.~(\ref{eq:rankine_m}) and (\ref{eq:rankine_p}), we obtain the linear relation 
between the pressure and specific volume $V=1/\rho$. Eliminating the velocity from Eq.~(\ref{eq:rankine_e}), on the other hand, we can derive the 
so-called Rankine-Hugoniot relation that gives the final pressure as a function of the initial specific volume and pressure as well as 
the final specific volume. Drawing these relations in the $P-V$ plane, we obtain the Rayleigh line from the former and the Hugoniot curve 
from the latter. 
The intersection of the line and curve gives the final state. 
If the initial and final states are in the same phase, 
the Hugoniot curve runs through not only the final state but also the initial state. 
This is the case for shock waves. 
If the initial 
phase is different from the final one, on the other hand, the Hugoniot curve does not pass through the initial state and it is different
from the Hugoniot curve for shock wave. What we are concerned with here is the relative positions of this Hugoniot curve for combustion and that for  shock wave.

\begin{figure}[t]
\includegraphics[width=14cm]{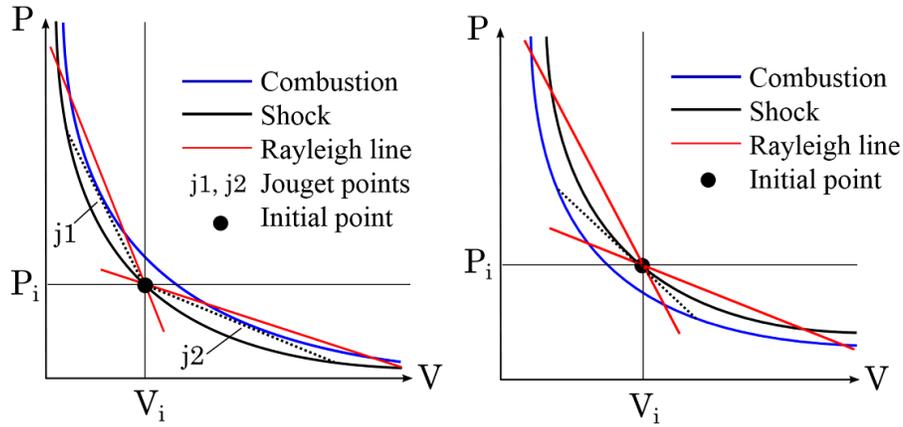}
\caption{The Hugoniot curves in the exothermic (left panel) and endothermic regime (right panel). The Hugoniot curves for shock wave are expressed as black curves.}
\label{fig_pvcurve}
\end{figure}

Since the Hugoniot curves for terrestrial combustions always run above the corresponding ones for shock waves in the $P-V$ plane, combustion theory
normally deals only with this case.  
Then the Rayleigh line intersects the Hugoniot curve at two points in general (except for the so-called Juget points, at which the two points coincide with each other). 
Depending on the slope of the Rayleigh line, or equivalently the flow velocity ahead of the combustion front, the two intersections lie either to the left or to the right of the point corresponding to the 
initial state. To be more precise, these four combustions are characterized by the ratios of the flow velocity to the sound speed, $c_{s}$, 
or the Mach numbers, for the initial and final states. If the relation $v_{i} > c_{si}$ holds, the combustion is referred to as "detonation", whereas 
it is called "deflagration" if the opposite inequality is satisfied. If, in addition, the relation $v_{f} < c_{sf}$ ($v_{f} > c_{sf}$) holds, 
it is said to be strong (weak). We hence have got four combustion modes: strong/weak detonation/deflagration. Since 
combustions are normally exothermic chemical reactions, they are referred to as the combustions in the "exothermic" regime.

It should be noted, however, that there is no a priori reason for not finding the opposite regime, which we refer to as the "endothermic" regime. 
In fact, as we will show later in this paper, the non-equilibrium phase transition from HM to 3QM is more likely than not to occur in the 
latter regime. As a matter of fact, we are not the first to point this out. Lugones et al.\cite{lugones94} gave a simple criterion of the 
exothermic combustion for the bag model and demonstrated that the endothermic regime is indeed obtained for the combination of 
the Walecka EOS for HM and the MIT bag model for 3QM. Simply put, the endothermic combustion obtains if HM is stiff enough. As shown 
shortly, this is indeed the case over a wide range of the bag constant and the strong coupling constant
 also for the baryonic EOS based on the relativistic mean field theory
that we employ in this paper. In the endothermic combustion, the Rayleigh line again intersects the Hugoniot curve at two points. Unlike in 
the exothermic case, however, there is no Juget point and one of the two intersections lies to the left of the initial point and the other sits on
the opposite side. There is no gap between detonation and deflagration in the fluid velocity ahead of the combustion front. These features are
summarized in a schematic picture in Fig.~\ref{fig_pvcurve}.
 
\begin{figure}[t]
\includegraphics[width=8cm]{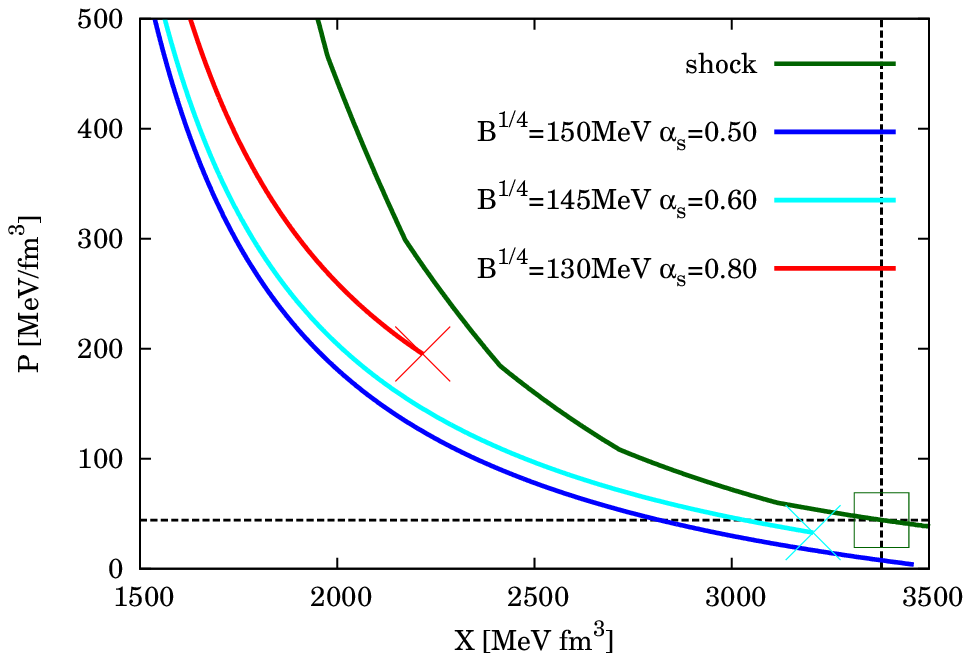}
\includegraphics[width=8cm]{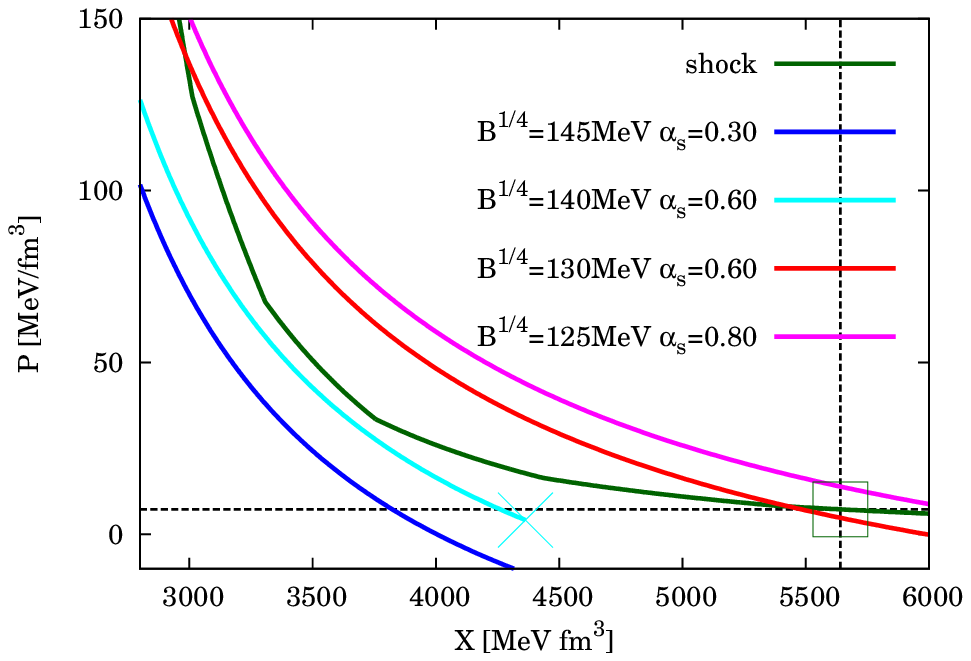}
\caption{The Hugoniot curves for  various  values of the bag  and strong coupling constants.
The initial states of HM are 
  NS matter in  $\beta$-equilibrium without neutrinos at $T =$ 0 MeV and  $\rho_i =$ 5.6 $\times 10^{14}$~g/cm$^3$  (left panel)  and
PNS matter at $T =$ 10 MeV, $Y_p=0.3$ and  $\rho_i =$ 3.0$\times 10^{14}$~g/cm$^3$  (right panel), respectively,
 which are indicated by the squares.
 Crosses mark the points, at which zero temperature is obtained.
}\label{fig_hugoniot1}
\end{figure}

 Some of the actual Hugoniot curves are given in  Fig.~\ref{fig_hugoniot1}, 
where the initial HM is assumed to be either cold neutron star (NS) matter
 with  the temperature $T=$~0 MeV and  the density $\rho_i =$ 5.6 $\times 10^{14}$~g/cm$^3$  or proto-neutron star (PNS) matter with $T=$10 MeV, $\rho_i =$ 3.0 $\times 10^{14}$~g/cm$^3$ and the lepton fraction $Y_{lep}$=0.3.
In both cases HM is assumed to be in  $\beta$-equilibrium either without (NS matter) or with (PNS matter) neutrinos.
Different colors correspond to different values of the bag constant,  $B^{1/4}$, and  the strong coupling constant,  $\alpha_s$.  
Note that the Hugoniot curves  in the figure are obtained from the relativistic version of 
Eqs.~(\ref{eq:rankine_m})-(\ref{eq:rankine_e})  and  $X = (h \rho)/n_B^2$  
 is reduced to the specific volume in the non-relativistic limit \cite{steinhardt}.
 As $B^{1/4}$ becomes larger,  the Hugoniot curve for combustion shifts downwards  due to the lower pressures for fixed
values of $n_B$ or $X$.  
In contrast,  $\alpha_s$ has little impact on  the position of Hugoniot curves, since the variations of $P$ and $X$  with $\alpha_s$ in the Rankine-Hugoniot relation are  mostly canceled out.
Both  $B^{1/4}$ and $\alpha$  give positive non-kinetic contributions to  internal energy
and, as a result, tend to lower temperatures and increase latent energies.
These features  are explained more in detail in Sec.~\ref{modeleos}. 
For comparison, the Hugoniot curves for shock wave in HM are also shown in the figure.

It is evident in the same figure that in most cases, 
the Hugoniot curves for combustion run below the corresponding Hugoniot curves for shock wave.
In the model  with $B^{1/4}$=130 MeV  and $\alpha_s=0.6$ for the PNS matter (right panel), the Hugoniot curve for combustion 
 intersects that for shock wave. Even in that case, however, the former is still lower than the latter near the initial point.
The  Hugoniot curve for combustion  runs above that for  shock wave 
for small bag constants as demonstrated by the model with $B^{1/4}$=125 MeV for the PNS matter (right panel).
It is interesting to point out that in  some cases  such as the model with $B^{1/4}$=140 MeV  and $\alpha_s=0.6$,
the Hugoniot curves cannot be extended to low pressures, since the temperature would become negative to obtain the Hugoniot relation, because of large latent energies.
The end points of the curves, which are marked with a cross in the figure, correspond to the states with $T=0$ MeV.
It is incidentally mentioned that these end points occur at higher pressures for larger $\alpha_s$, as observed for
 the models with $B^{1/4}=130$ and  $\alpha_s=0.8$ for NS matter (left panel).
These are the real end points, below which we have no steady solution that satisfies the conservation laws. 
What happens then will be an interesting topic that warrants further investigations.
Note in passing that the final state is not always pure 3QM as assumed tacitly in the above discussions.
 In some case,  HM survives even in the final state in chemical equilibrium as explained later.

As mentioned earlier, in the studies done so far, the endothermic regime is simply discarded for the very reason that it is the endothermic regime. The idea behind 
this seems to be that endothermic reactions do not occur at zero temperature. It should be stressed, however, that the term "endothermic regime" 
is misleading and indeed does not necessarily correspond to the reactions being endothermic. As explained above, the regimes are classified by the 
relative positions of the two Hugoniot curves, one for combustion and the other for shock wave: if the Hugoniot curve for combustion runs above 
(below) the one for shock wave at the specific volume for the initial state,
the combustion is referred to as exothermic (endothermic). 
The heat release or absorption, however, is given by $d'Q = dU - PdV$ (in the absence of mass actions) 
according to the first law of thermodynamics.
 It should be noted that in the phase transition we are concerned with in this paper, neither the 
specific volume nor pressure remains constant. The terminology is simply based on the our empirical knowledge that the Hugoniot curves for 
terrestrial exothermic combustions always run above the corresponding Hugoniot curves for shock wave. 

We emphasize here that it is not whether the phase transition is exothermic or endothermic in the conventional sense above that matters.
 What is more important is whether 
the free energy is minimum or not. If not, the reaction can proceed toward the free energy minimum irrespective of it being exothermic or not
in the conventional sense. This is exactly what we have at 
hand. Indeed,  the free energy is always lower for 3QM at the end points than for HM at the base points of arrows $a$ and $b$ both for cases (A) and (B)  in Fig.~\ref{fig_scenario_energytoy}. We hence contend that the phase transition should occur there 
even if they are in the endothermic regime. The conversion will be terminated only when $P_{c3}$ in case (B) is reached by the conversion front.

In this subsection we have ignored the width of conversion front and discussed only the possible asymptotic states. In the following, we will see 
what structures the conversion fronts have according to the scenarios described above. 
In the rest of this paper, we employ a non-relativistic formulation for simplicity.
 This is certainly not a good approximation. 
The reasons why we make this choice are
 that the non-relativistic dynamics and
 Hugoniot relations are no doubt easier to
 understand intuitively, that there is a subtle problem
 of causality violation in the naive relativistic
 extension of our model and that the results are
 qualitatively unchanged according to our preliminary explorations.
 Fully relativistic and more quantitative analysis will be reported in the sequel to this paper. 
In order to further expedite the understanding, we first employ 
a simplified model, which captures only the essential ingredients of the more realistic model discussed in Section~\ref{model}.  

\section{Toy Model \label{toy}}

In the following, we turn our attention to the structures of the conversion region connecting the initial and final asymptotic states
discussed in the previous section. We assume the plane symmetry and consider one dimensional stationary profiles of matter flows 
that undergo the phase transition from HM to 3QM. The assumption of plane-symmetry and stationarity is well justified, since the width of the 
conversion region is much smaller than the typical macroscopic length scale and the time, during which matter stays in this region, 
is much shorter than the time scale, on which the initial hadronic state is changed either by the propagation of the conversion front 
in the (proto) neutron star or by the adjustment of  (proto) neutron-star configuration to the appearance of quark phase. In this section, we introduce 
a toy model that will facilitate our analysis and understanding of the main results presented in Section~\ref{model}. The simplification 
is mainly concerning EOS's. As shown shortly, it is indeed a very crude approximation to reality. However, the qualitative behavior of its 
results still captures the essence of the
more realistic models introduced in the next section. There is also an advantage that we can freely change the behavior of Hugoniot 
curves, particularly the regime of combustion. We hence believe that this simplified model is worth presenting here.  
 
\subsection{basic equations \& simplified EOS's\label{toyeq}}

The basic equations to describe the stationary structures of the conversion region are the conservation equations of mass, momentum
and energy in the front-rest frame, which are unchanged in the more realistic models introduced in the next section and given by 
\begin{eqnarray}
\rho v &=& \rho _i v_i \ (= \rho _f v_f), \label{eq:eq4}\\
P+\rho v^2 - \nu \frac{dv}{dx} &=& P_i +\rho _i v_i ^2 \ (= P_f +\rho _f v_f ^2), \label{eq:eq5}\\
h + \frac{1}{2} v^2 - \frac{\nu}{\rho} \frac{dv}{dx} &=& h_i + \frac{1}{2}v_i ^2 \ (=  h_f + \frac{1}{2}v_f ^2), \label{eq:eq6}
\end{eqnarray}
where plane symmetry is assumed and an $x$ coordinate is introduced; the initial HM is assumed to be located at $x = -\infty$ and the 
final 3QM is assumed to be realized at $x = +\infty$; $\rho$, $v$, $P$, and $h$ are the baryon density, fluid velocity, 
pressure and specific enthalpy, respectively; the subscripts $i$ and $f$ stand for the initial and final states as before; the viscous dissipations 
are introduced to deal with shock waves, in which $\nu$ is the viscosity. 
These equations are complemented by another equation that 
gives the spatial distribution of strangeness,
\begin{eqnarray}
v \frac{ df_s } { dx }  = \frac{f_{s,f}-f_s}{\tau }  \label{eq:eq7}, 
\end{eqnarray}
where $f_s$ is the fraction of strangeness and $f_{s,f}$ is its asymptotic value in the final state;  $\tau$ gives the time scale of $\beta$-equilibration; they are varied rather arbitrarily in the toy model to see the
dependence of solutions on these parameters. Divided by $f_{s,f}$, the above equation is rewritten as  
\begin{eqnarray}
v \frac{d\bar{f}_s}{dx}  = \frac{1-\bar{f}_s}{\tau}, \label{eq:eq8}
\end{eqnarray}
for $\bar{f}_s=f_s/f_{s,f}$.

 The $\beta$-equilibration occurs once the deconfinement to 2QM is allowed. Although in our realistic models, this point is determined by 
the condition for the phase equilibrium between HM and 2QM, in the toy model we set it arbitrarily. The condition is usually met inside 
the shock wave (see the left panel of Fig.~\ref{fig_structure}), in which the density increases by compression. This is the reason why we do not treat 
the shock wave as a discontinuity but calculate its profile employing the viscous dissipation term. Note that the width of the shock
wave is of the same order as that of the deconfinement region and much smaller than the length scale of the conversion region of our interest.
In principle, what happens inside the shock wave cannot be described by the hydrodynamical approximation. It is a common practice, however, 
and it is also known that the qualitative behavior is reproduced. We thus follow this practice, adjusting the viscosity so that 
the shock wave should be much narrower than the conversion region but could be still resolved in numerical calculations.

We employ the so-called $\gamma$-law EOS both for HM and QM, knowing that this is certainly an oversimplification:
\begin{eqnarray}
P_{HM}=(\gamma -1)\rho \epsilon, \label{eq:eq9}\\
P_{QM}=(\gamma -1)\rho (\epsilon +e), \label{eq:eq10}
\end{eqnarray}
where the upper equation is for HM and the lower for QM; $\gamma $, $\rho $ and $\epsilon $ are the adiabatic index, baryon density and 
specific internal energy, respectively. The EOS for QM is different from that for HM in that the former includes an extra constant term, $e$, 
in the specific internal energy, which is utilized to control the regime of combustion; with a positive $e$, we have an exothermic 
combustion and vice versa in the conventional sense. 

In the conversion region, QM has strangeness fractions that are different from the asymptotic value. In this section, we assume
for simplicity that these states are also described by the $\gamma$-law EOS,
\begin{eqnarray}
P = (\gamma -1)\rho (\epsilon +\bar{f}_s e), \label{eq:eq11}
\end{eqnarray} 
where we multiply the extra energy, $e$, with the fraction of strange quark, $\bar{f}_s$, introduced above, thus interpolating the intermediate
2QM ($\bar{f}_s=0$) and final 3QM ($\bar{f}_s=1$) very crudely. These treatments will be much sophisticated in Section~\ref{model}. Since we are 
interested in the qualitative features of the conversion region in this section, this level of approximation is sufficient. 

We normalize all quantities by adopting an appropriate density, pressure and time, for which we normally take the initial density, 
pressure and weak interaction time scale. Then the parameters that characterize the system are the dimensionless viscosity, $\bar{\nu}=\nu / (\rho_i c_{si}^2 \tau) $  and extra energy, $\bar{e}=e/c_{si}^2 $.

The toy model can be extended to the diffusion-induced conversion by adding the second spatial derivative in Eq.~(\ref{eq:eq7}), which describes the diffusion flux of strangeness. On the other hand, the viscous term in Eqs. (\ref{eq:eq5}) and (\ref{eq:eq6}) can be neglected in that conversion. Detailed analyses of the modified toy model as well as the more realistic one, which corresponds to what is introduced in the next section of this paper, will be postponed to the sequel to this paper.

\subsection{results}
In the following we show the solutions to the equations given above (Eqs.~(\ref{eq:eq4})-(\ref{eq:eq6}) and (\ref{eq:eq8}) together 
with Eqs.~(\ref{eq:eq9})-(\ref{eq:eq11})). The exothermic ($e>0$) and endothermic ($e<0$) cases are
discussed in turn separately. 

\subsubsection{exothermic case ($e>0$) \label{toyex}}
We first consider the exothermic case with $e>0$, i.e., the ordinary combustion as observed on earth. 
The Hugoniot curve for combustion then runs above the one for shock wave in the $P-V$ diagram. 
We suppose here that a shock wave traverses HM; the shock compression increases the density 
and pressure and at some point inside the shock wave 2QM is favored in terms of free-energy ($P_{c2}$ in Fig.~\ref{fig_scenario_energytoy}); then 
the phase transition to 2QM occurs spontaneously, followed by the $\beta$-equilibration to 3QM. 
The basic equations are reduced to the autonomous system for $\bar{f}_{s}$ and $v$:
\begin{eqnarray}
\frac{d\bar{f}_{s}}{dx}&=&\frac{1-\bar{f}_{s}}{v}, \label{eq:fs_sh}\\ 
\frac{dv}{dx}&=&\frac{1}{2 \bar{\nu} v}\left\{(\gamma+1)M_i (v-M_i)\left(v-\frac{2+(\gamma -1)M_i^2}{(\gamma +1)M_i}\right)+2(\gamma -1)M_i \bar{e} \bar{f}_{s}\right\}, \label{eq:vl_sh}
\end{eqnarray}
where $M_i$ is a Mach number for the initial state and all quantities are normalized as mentioned earlier. The right hand side of
Eq.~(\ref{eq:fs_sh}) is turned on only for the QM region. The critical point, at which HM converts itself to 2QM spontaneously, is a free parameter
specified by the density, $\rho_{c}$, in this toy model. In more realistic models we employ in the next section, the critical point is determined 
by the condition for the phase equilibrium between HM and 2QM, which is hence consistent with the EOS's adopted there. Our concern here is whether 
strong detonation is the only solution or weak detonation is also realized. Other possibilities are unlikely, since decompression is needed.

\begin{figure}[t]
\includegraphics[width=8cm]{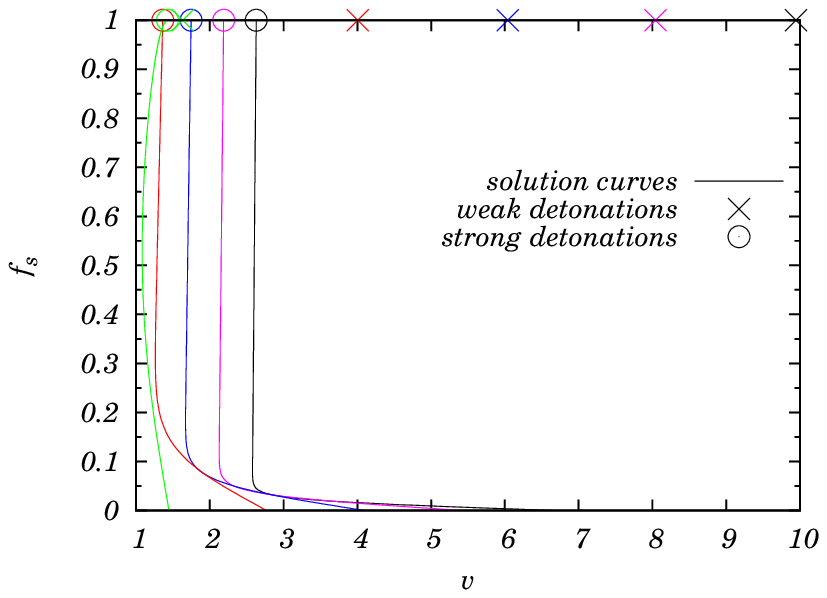}
\includegraphics[width=8cm]{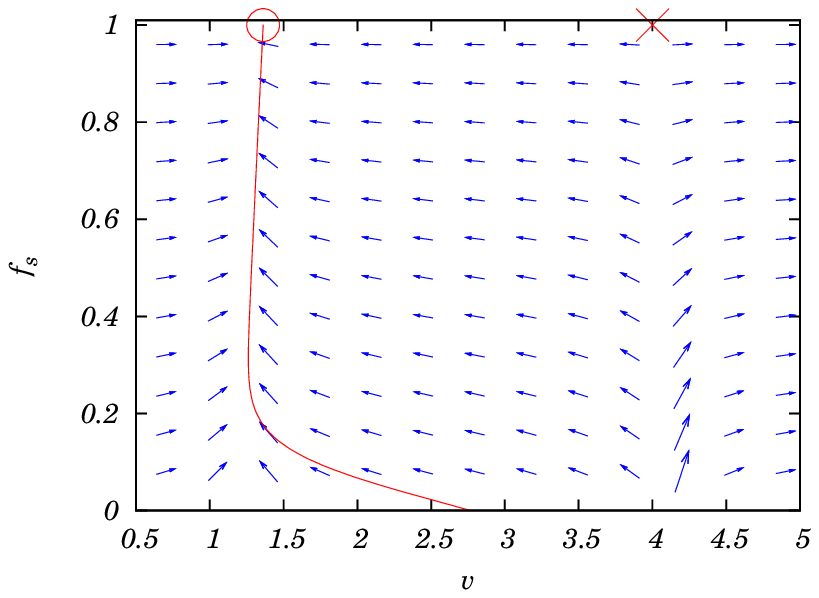}
\caption{The phase diagrams for the toy model of the shock-induced conversion in the exothermic regime.
 Left panel shows some solution curves for different initial velocities drawn in different colors.
 Right panel displays the vector field  ($dv/dx$,  $d\bar{f}_{s}/dx$)
for a given initial Mach number and the corresponding solution curve.
The final states corresponding to strong (weak) detonation are marked by circles (crosses).}
\label{fig_toymodel_ph_sh}
\end{figure}

 For a given pair of $(V, P)$, we obtain a 
one-parameter family of solutions and 
integral curves in the phase space spanned in this case by $v$ and $\bar{f}_s$. Some representative solutions are given in the left panel of  Fig.~\ref{fig_toymodel_ph_sh}.
The initial states correspond to the points on the $v$-axis ($\bar{f}_{s}=0$) in this figure whereas the final states are the points on the line, 
$\bar{f}_{s}=1$. The latter is divided into two regions, one for strong detonation and the other for weak detonation, each of which is marked with circle and cross, respectively.
 It is evident that strong detonation is always obtained,
 ie, all integral lines go into the portion of strong detonation. 
The right panel of  Fig.~\ref{fig_toymodel_ph_sh} shows that 
the vector field ($dv/dx$, $d\bar{f}_{s}/dx$), which are derived from Eqs.~(\ref{eq:fs_sh}) and (\ref{eq:vl_sh}) for some initial Mach number.
We can confirm that all but one integral lines converge to the asymptotic state corresponding to the strong detonation and 
  weak detonation is unlikely to be realized.
Note that the vector field depends on the initial Mach number (see Eq.~(\ref{eq:vl_sh})) and is different for each initial state.

\begin{figure}[t]
\includegraphics[width=8cm]{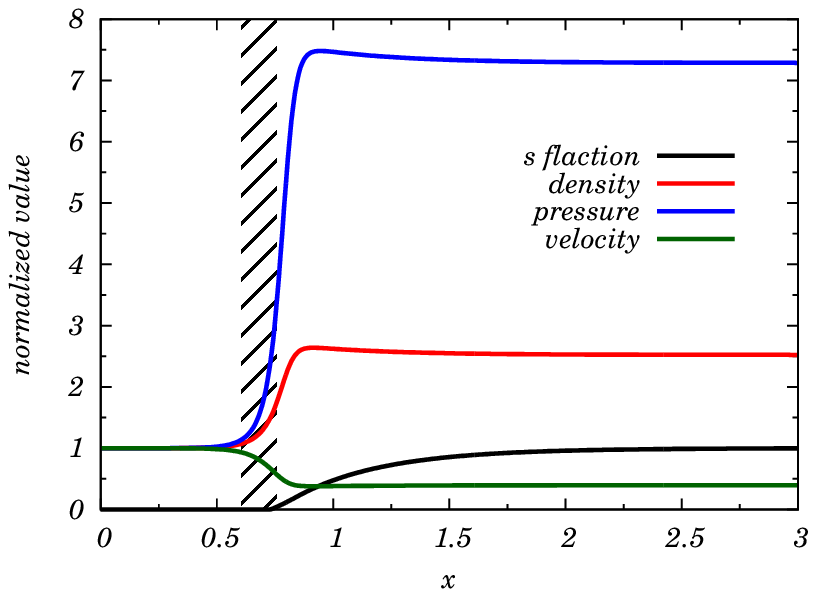}
\includegraphics[width=8cm]{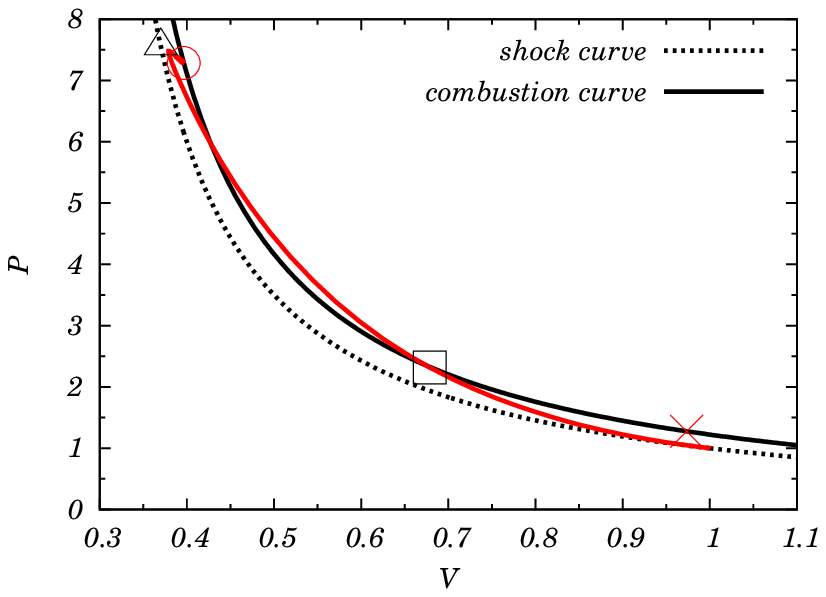}
\caption{The typical strong-detonation solution  for the toy model of the shock-induced conversion in
the exothermic regime displayed  as functions of position (left panel) and as a trajectory in the P-V diagram (right panel). 
With a non-vanishing viscosity, the shock is no longer a discontinuity but has a finite width,
which is hatched in the left panel. In the right panel, the Hugoniot curves for combustion and shock
wave are drawn as solid and dotted lines, respectively. The actual evolution is expressed with the
red curve, which is not a straight line indeed. The final state corresponding to strong (weak) detonation is marked by circle (cross).
The critical point with $\rho=\rho_c$  and final state for the shock transition without combustion are also indicated by square and triangle, respectively.}
\label{fig_toymodel_sol}
\end{figure}

A typical solution of strong detonation is displayed in Fig.~\ref{fig_toymodel_sol}. In the left panel, we show various quantities as a function
of position. The shock wave is located at $x \sim 0.75$. With a non-vanishing viscosity the shock wave is no longer a discontinuity but 
is smeared over a finite region, in which the velocity and pressure change very rapidly. 
The fact that this is actually a strong detonation is best demonstrated in the right panel, in which we show the solution together with the Hugoniot curves both for shock wave and combustion in the $P-V$ diagram. The strong and weak detonations
are marked with circle and cross, respectively,  on the Hugoniot curve for combustion whereas the final state that would be connected by the shock transition without 
combustion is also indicated by a triangle on the Hugoniot curve for shock wave. 
Note that the trajectory is not a straight line owing to the viscous term and does not go through the final state for weak detonation. 
As expected, the specific volume and pressure initially move toward the final state for the shock transition. 
The evolution is 
redirected to the final state corresponding to strong detonation much later than the point marked with square, at which the critical density, $\rho_c$, for the deconfinement from HM to 2QM is reached.

One might think that if the critical density were closer to the density of the final state corresponding to  weak detonation, the resultant mode might be different.
This does not seem the case. In fact, we studied a wide range of $\rho_c$,
 to always obtain strong detonation.
Note also that the viscosity ($\bar{\nu}=1$)
adopted in these calculations is way too large. As a matter of fact, with a realistic value, the shock width would be smaller by many orders of magnitude 
than the size of the conversion region that is determined by weak interactions as mentioned earlier. This is the main
reason why we take such an overly exaggerated value here and in the next section as well. We confirm at least that the results presented above
are not changed if we take much smaller but still treatable values of viscosity. It is also mentioned incidentally that in this toy model the conversion 
from HM to 2QM is assumed to be completed instantaneously but this will be improved in the more realistic model, which takes the so-called mixed phase into account,  in the next section.

\begin{figure}[t]
\includegraphics[width=8cm]{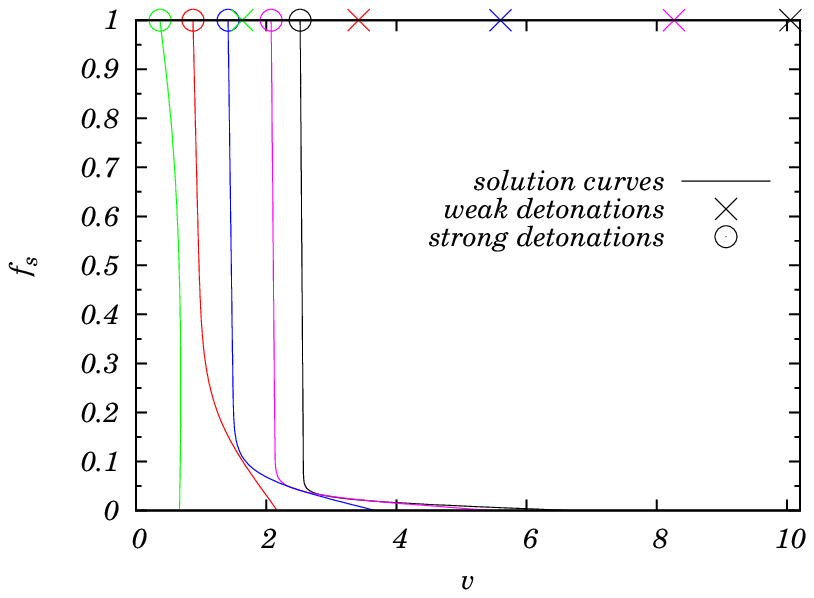}
\includegraphics[width=8cm]{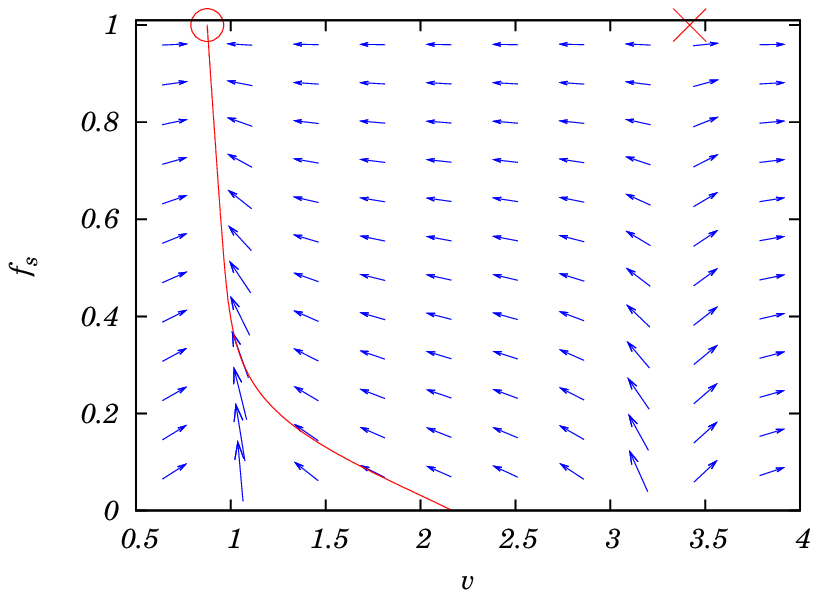}
\caption{The phase diagrams for the toy model of the shock-induced conversion in the endothermic regime.
Lines and symbols are the same as in Fig.~\ref{fig_toymodel_ph_sh}.}
\label{fig_toymodel_en_ph}
\end{figure}

\begin{figure}[t]
\includegraphics[width=8cm]{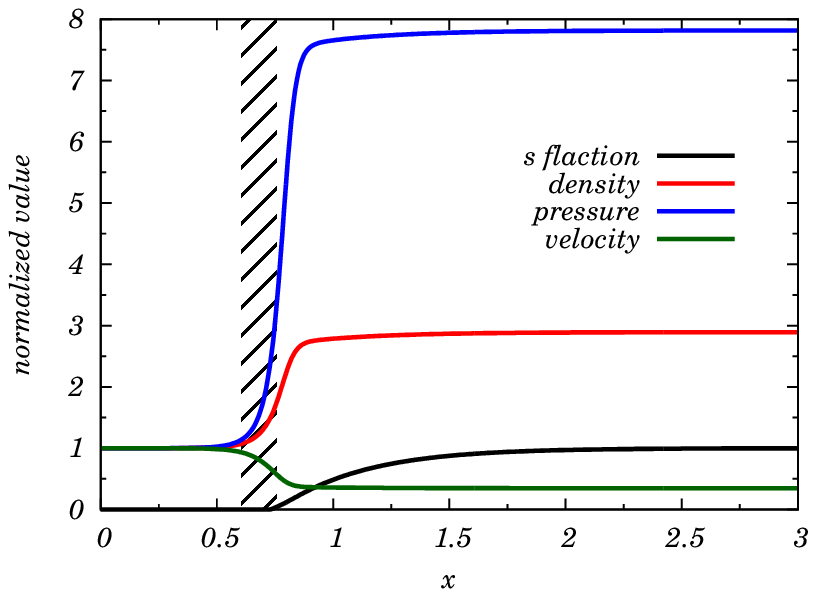}
\includegraphics[width=8cm]{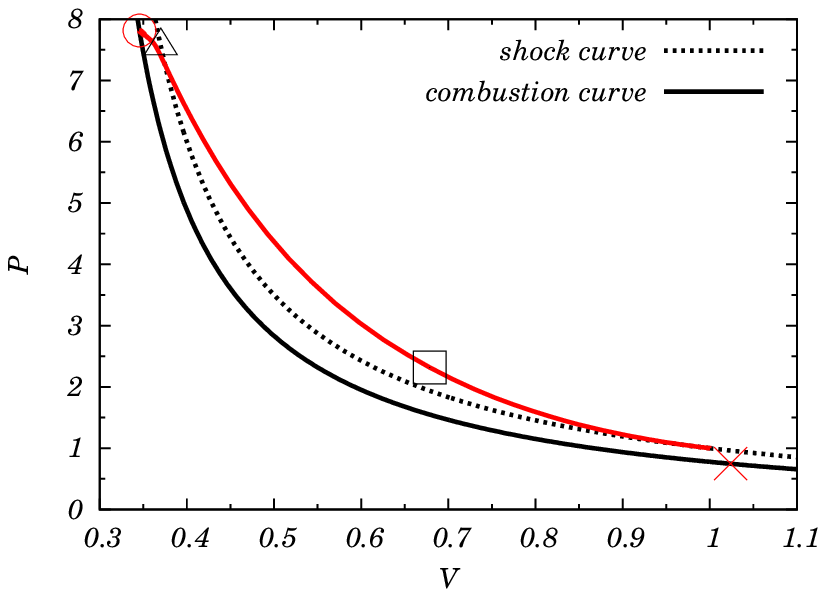}
\caption{ The typical strong-detonation solution for the toy model of the shock-induced conversion in the endothermic regime.  
Lines and symbols are the same as in Fig.~\ref{fig_toymodel_sol}
}   
\label{fig_toymodel_en_sol}
\end{figure}

\subsubsection{endothermic case ($e<0$)\label{toyen}}
Now we proceed to the endothermic case ($e<0$), the exotic combustion, to which we have no terrestrial counter part~\cite{williams, barton1973}.
The Hugoniot curve for combustion runs below the one for shock wave in this case. Like the exothermic case, there are two states that satisfy
the Rankine-Hugoniot jump conditions for a given pair of $(V, P)$ and a velocity $v$. Unlike the ordinary combustion, however, we always find one of 
them to the left and the other to the right of the initial state in the $P-V$ diagram. 
These combustions are classified by the same scheme as for the exothermic case: detonation is a combustion mode, for which the initial state is supersonic in the front-rest frame, whereas deflagration 
is a combustion with a subsonic initial velocity; if the final state is subsonic, the combustion is either strong detonation or weak deflagration; 
on the other hand, it is called either weak detonation or strong deflagration if the flow in the final state is supersonic. One interesting feature in
the endothermic combustion is that there is no Jouget point and the detonation branch is connected with the deflagration branch without a gap in the
initial velocity. 

In the same way for the exothermic case, weak detonation is unlikely to be realized as shown in Fig.~\ref{fig_toymodel_en_ph}. 
Hence strong detonation is expected to be the only solution.
In fact,  matter is decompressed in weak detonation whereas it is compressed in the shock wave.
A typical  strong-detonation solution is  presented in Fig.~\ref{fig_toymodel_en_sol}.
 For the comparison with the exothermic counterpart (Fig.~\ref{fig_toymodel_sol}),
it is found that the distributions of various quantities are 
not much different between the two cases except that matter is further compressed
in the endothermic combustion after the transition
from HM to 2QM commences in the shock wave. This is more evident in the right panel, in which we show the trajectory together with the
Hugoniot curves for combustion as well as for shock wave in the $P-V$ diagram.
 In these calculations, we again take the viscosity $\bar{\nu}$ that is too
large for the same reason as given earlier for the exothermic case.
 It is confirmed that the results are unchanged if we take smaller (but still affordable) values of $\bar{\nu}$.

\section{Formulations of Realistic Models \label{model}}
In the following, we elaborate on the micro physics neglected in the toy model and study with this more realistic model what we have found so far with the 
overly simplified model. 
 
\subsection{EOS's for HM and QM \label{modeleos}}
The EOS for ideal gas used in the toy model is certainly inappropriate both for HM and
QM and is replaced by more realistic (albeit phenomenological) ones, which will be described
below in turn.
We employ the Shen's EOS~\cite{shen11} for HM, which is based on the relativistic mean field theory, in which nuclear interactions are described by exchanges of mesons.
The Lagrangian
adopted in the Shen's EOS is the following:
\begin{eqnarray}
{\cal L}_{RMF} & = & \bar{\psi}\left[i\gamma_{\mu}\partial^{\mu} -M
-g_{\sigma}\sigma-g_{\omega}\gamma_{\mu}\omega^{\mu}
-g_{\rho}\gamma_{\mu}\tau_a\rho^{a\mu}
\right]\psi  \\ \nonumber
 && +\frac{1}{2}\partial_{\mu}\sigma\partial^{\mu}\sigma
-\frac{1}{2}m^2_{\sigma}\sigma^2-\frac{1}{3}g_{2}\sigma^{3}
-\frac{1}{4}g_{3}\sigma^{4} \\ \nonumber
 && -\frac{1}{4}W_{\mu\nu}W^{\mu\nu}
+\frac{1}{2}m^2_{\omega}\omega_{\mu}\omega^{\mu}
+\frac{1}{4}c_{3}\left(\omega_{\mu}\omega^{\mu}\right)^2   \\ \nonumber
 && -\frac{1}{4}R^a_{\mu\nu}R^{a\mu\nu}
+\frac{1}{2}m^2_{\rho}\rho^a_{\mu}\rho^{a\mu}. 
\end{eqnarray}
The notation is the same as in \cite{shen11}: $\psi$, $\sigma$, $\omega$ and $\rho$ denote nucleons 
(proton and neutron), scalar-isoscalar meson, vector-isoscalar meson and vector-isovector meson, respectively, and 
$W_{\mu\nu} =\partial^{\mu}\omega^{\nu}- \partial^{\nu}\omega^{\mu} $ and 
$R^{a}_{\mu\nu}= \partial^{\mu}\rho^{a\nu}- \partial^{\nu}\rho^{a\mu} +g_{\rho}\epsilon^{abc}\rho^{b\mu}\rho^{c\nu} $. 
The nucleon-meson interactions are given by the Yukawa couplings and the isoscalar mesons ($\sigma$ and 
$\omega$) interact with themselves, which are expressed as the cubic and quartic terms. 
These self-interactions are tuned to reproduce more fundamental Br${\rm\ddot{u}}$ckner-Hartree-Fock theory.
In the mean field theory, the mesons are assumed 
to be classical and replaced by their thermal  ensemble averages whereas the Dirac equation for nucleons is quantized and the free energy is 
evaluated based on the energy spectrum of nucleons obtained this way. $M$ is the mass of nucleons and assumed to be $938$~MeV. 
They use  the TM1 parameter set, in which the masses of mesons, $m_{\sigma}$, $m_{\omega}$, $m_{\rho}$, and 
the coupling constants, $g_{\sigma}$, $g_{\omega}$, $g_{\rho}$, $g_2$, $g_3$, $c_3$, are determined so that not only the saturation of 
uniform nuclear matter but also the properties of finite nuclei could be best reproduced \citep{sugahara94}. 
This EOS is rather stiff, having the incompressibility of 281 MeV and the symmetry energy of 36.9 MeV, and the maximum mass of cold neutron star is  2.$2M_{\odot}$.

We adopt the MIT bag model for QM, which takes into account the confinement
and asymptotic freedom of quarks phenomenologically and describes QM as a collection of
freely moving quarks in the perturbative vacuum with a vacuum energy density given by the so-called bag constant.
The baryonic number density $n_B$, pressure $P$ and energy density $\epsilon $ are expressed as the sums over three flavors of quarks, which are shown by the subscripts
of $u$, $d$, $s$ for up, down and strange quarks, respectively, and the bag constant $B$:
\begin{eqnarray}
n_{B} &=& \frac{1}{3}\sum _{f=u,d,s} n_{f}, \label{eq:bag1}\\
P &=& \sum _{f=u,d,s} P_{f} - B, \label{eq:bag2}\\
\epsilon &=& \sum _{f} \epsilon _{f} + B. \label{eq:bag3}
\end{eqnarray}
The contributions of  each  flavor of quark to these quantities are given as
\begin{eqnarray}
P_f &=& \frac{g_f}{6\pi ^2} \int ^{\infty} _0 \frac{p_f^4}{\sqrt{p_f^2+m_f^2}} [F(E,\mu _f, T) + F(E,-\mu _f, T)] dp_f, \label{eq:pre_f} \\
\epsilon _f &=& \frac{g_f}{2\pi ^2} \int ^{\infty} _0 p_f^2\sqrt{p_f^2+m_f^2} [F(E,\mu _f, T) + F(E,-\mu _f, T)] dp_f, \\
n_f &=& \frac{g_f}{2\pi ^2} \int ^{\infty} _0 p_f^2 [F(E,\mu _f, T) + F(E,-\mu _f, T)], \\
\end{eqnarray}
where $g_f (= 6)$, $p_f$, $E_f$ and $\mu _f$ are the statistical weight, momentum, energy and chemical
potential, respectively, of each flavor of quark, and the Fermi-Dirac distribution is denoted
by $F(E_f,\mu _f, T) = 1/(e^{(E _f - \mu _f)/T)}+1)$ with $E = \sqrt{p_f^2 + m_f^2}$ and $m_f$ being the quark mass.
The masses of quarks are set to be $m_u=2.5$, $m_d=5.0$ and $m_s=100$ MeV
 \cite{nakamura10}.
The statistical weight
 $g_f$ is  a product of  
 spin (2) and color (3) degrees of freedom.
In these simplest expressions, the interactions of quarks are neglected except for the bag constant. 
Fahri et al. \cite{farhi84} derived the first-order corrections with respect to the strong coupling constant for massless quarks at finite temperatures.
We add them to 
the simplest expression (Eq. (\ref{eq:bag3})) for massive quarks following
 \cite{fischer2010} as
\begin{eqnarray}
P_f (\alpha _s) = P_f(0)- \left[ \frac{7}{60}T^{4}\pi ^{2}\frac{50\alpha_s}{21\pi } + \frac{2\alpha _s}{\pi }\left( \frac{1}{2}T^2 \mu_f^2 + \frac{\mu_f ^4}{4\pi ^2} \right) \right],
\label{eq:alpha}
\end{eqnarray} 
where $P_f(0)$ is the pressure given by   Eq.~(\ref{eq:pre_f}).
The number density $n_f$ and energy density $\epsilon _f$  can be obtained in a similar way. 
 
\begin{table}[htb]
  \begin{tabular}{|c|c||c|c|c|c|c|}
  \hline
   $B^{1/4}$ \ & \ $\alpha_s$ \ & \ $M_{max}$ [M$_{\odot}$] \ & \ 2QM [MeV]  \  & \  3QM [MeV] &\   $n_{c}/n_{0}$  of NS matter   &\ $n_{c}/n_{0}$  of PNS matter \\  \hline \hline
    145 & 0.40 & 1.90 & 1017 &  922 & 1.17& 2.40  \\ \hline
    140 & 0.40 & 2.04 & 970&  892 & 0.75& 1.35  \\ 
    140 & 0.60 & 2.02 &1016  & 939  & 1.51& 4.37  \\ \hline
   135 & 0.60 & 2.17& 993 & 908  & 0.95 & 4.09  \\ 
   135 & 0.80 & 2.14& 1051 & 967  & 6.66 &  7.46  \\ \hline
    130 & 0.80 & 2.30 & 1001 & 934  & 6.38 &  7.30  \\ \hline
   125 & 0.80 & 2.47 & 962 & 907 & 0.74 &  1.21 \\ \hline
   \end{tabular}
\caption{Summary of models: the values of  the maximum mass of pure quark star, free energies of  2QM and 3QM at zero pressure and critical densities NS matter and  of PNS matter normalized by the nuclear density for different combinations of the bag and coupling constants.}
\label{table1}
\end{table}
 
\begin{figure}[t]
\includegraphics[width=10cm]{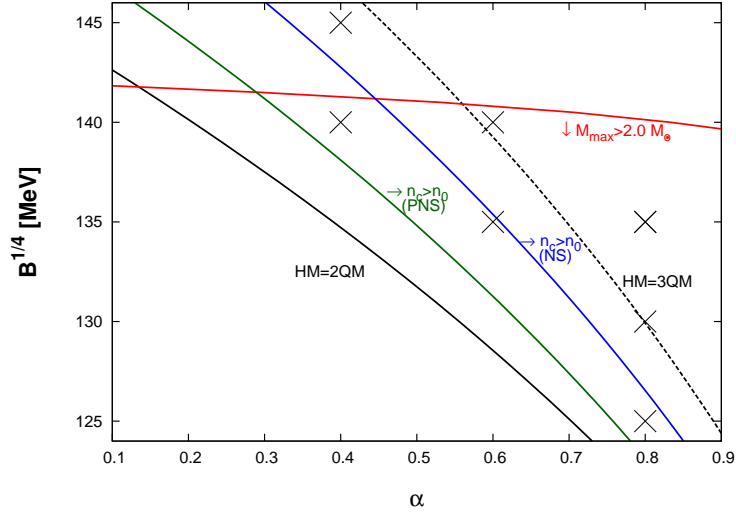}
\caption{
Some constraints on the values of the bag constant and the strong coupling constant.
The black solid and dashed lines show the pair, for which the energy per baryon of HM coincides with that of 2QM and with that of 3QM, respectively. 
The domain to the left of the former line is excluded, since HM would be unstable to the deconfinement to 2QM. 
The red curve is the critical line, above which the maximum mass of quark star would become smaller than  $2M_{\odot}$, the largest pulsar mass  observed so far.
 The green and blue solid lines indicate the pairs, for which the critical density of the spontaneous transition from HM to 2QM occurs at the nuclear saturation density. See the text for more details.
 The crosses correspond to the models investigated in this paper and listed in Table~\ref{table1}.}
\label{alphabag}
\end{figure}

\renewcommand\thefootnote{\alph{footnote}}
We summarize  in Table~\ref{table1} the combinations of the bag constant and the strong coupling constant we adopt in this paper and the resultant properties of 2- and 3QM's as well as of quark stars.
Note that all models in Table~\ref{table1} satisfy the requirement that 2QM in vacuum
should have a larger energy per baryon than that of HM  $\sim 934$ MeV including surface effects \cite{weissenborn}.
On the other hand, the  SQM hypothesis is true  for the four models,
 in which the  energies per baryon of 3QM are smaller than  $\sim930$ MeV.
 In Fig.~\ref{alphabag}, we indicate the parameter values of our choice in the $B-\alpha_s$ plane together with some critical lines. 
For example, the domain to the left of the black solid line labeled with "2QM$=$HM" is excluded, since 2QM would be more stable than HM there. The SQM hypothesis holds, on the other hand, for the parameter sets located to the left of the black dashed line with the label of "3QM$=$HM". In this paper we do not stick to this hypothesis and explore the regions both to the left and right of this critical line. The maximum mass of cold quark star would be smaller than  2 M$_{\odot}$, the mass of the most massive pulsar observed so far, above the red line.
 One of our models violates this condition indeed. In this case, quark star cannot exist as a stable object$^{\rm \footnotemark[1]}$.
 \footnotetext[1]{Quark star may still exist as a transient object if it is rotating sufficiently fast.}  
The critical density, at which HM converts itself  to 2QM spontaneously,
plays a key role in our scenario.
 Table~\ref{table1} gives the values of the critical density for the combinations of the bag constant and strong coupling constant we adopt in this paper. Both the transitions from the cold neutron star and from the hot and proton-rich proto-neutron star are considered. It is found 
 that the models with larger $B^{1/4}$ and/or $\alpha_s$
 have larger critical densities. 
This is because  2QM  with larger $B^{1/4}$ and/or $\alpha_s$ has smaller pressure for a given chemical potential,
as can be understood from Eqs.~(\ref{eq:bag1}-\ref{eq:bag3}) and (\ref{eq:alpha}). 

Figure~\ref{alphabag} shows
as blue and green solid lines for NS and PNS, respectively, 
the boundaries of the pairs of $B^{1/4}$ and $\alpha_s$, for which the critical density is larger than the nuclear saturation density
 $n_0=0.17$ fm$^{-3}$. 
We suppose that the conversion from HM to QM occurs at supra-nuclear densities and choose pairs of $B^{1/4}$ and $\alpha_s$ only above these lines in this paper.
 Recent observations of masses of compact stars also give a strong constraint on their EOS.
In particular, the observations of the pulsars PSR J1614-2230 and PSR J0348+0432
indicate that their masses are $1.97\pm$ 0.04~M$_{\odot}$ and $2.01\pm$ 0.04~M$_{\odot}$, respectively \cite{demorest,antoniadis13}. 
As mentioned earlier, we consider both the models that satisfy this constraint and that do not. In the latter case, quark star does not exist as a stable compact object and may be realized as a transient \cite{drago14}.

\subsection{Asymptotic states \label{modelstate}}%
In the following three subsections, we explain our descriptions of various regions in Fig.~\ref{fig_structure} in turn. We start with the asymptotic regions or the states, from one of which the conversion begins and with the other of which it ends. The former is called fuel in combustion and the latter is referred to as ash. 
The fuel of our interest is composed of neutrons and protons as the hadronic component$^{\rm \footnotemark[2]}$
\footnotetext[2]{Hyperons are assumed to be absent in this paper.}
and  electrons and neutrinos as the leptonic component.
They are assumed to be in   $\beta$-equilibrium,
\begin{eqnarray}
\mu _p +\mu _e = \mu _n +\mu _{\nu_e}.  \label{eq:betaf}
\end{eqnarray}
In the case of PNS ($T\sim 10$ MeV), neutrinos are assumed to be trapped and equilibrated with matter whereas they are assumed to be absent in the case of cold NS and we set $\mu_{\nu_e}=0$.
In addition, the fuel is charge neutral:
\begin{eqnarray}
n_{p}=n_{e},   \label{eq:chargef}
\end{eqnarray}
where $n_p$ and $n_e$ are the number densities of proton and electron, 
 respectively.
The lepton fraction $Y_{lep}$ is assumed to be
 $Y_{lep}=0.3$ everywhere for the PNS case. 
The electron fraction in the NS matter is determined
 from the conditions
of  $\beta$-equilibrium,  Eq.~(\ref{eq:betaf}), without neutrinos and of charge neutrality, Eq.~(\ref{eq:chargef}). 

On the other hand, the thermodynamic state of the ash or 3QM in  $\beta$-equilibrium
is derived from
\begin{eqnarray*}
\mu _{up} +\mu _e &=& \mu _{dn} +\mu _{\nu_e}, \\
\mu _{up} +\mu _e &=& \mu _{sg} +\mu _{\nu_e}, \\
\mu _{dn} &=& \mu _{sg}, 
\end{eqnarray*}
where each suffix: $up$, $dn$ and $sg$ stands  for up-, down- and strange-quarks, respectively. 
Neutrinos appear via weak interactions and we can not neglect them in the ash regardless of 
 whether they exist in the fuel or not. 
Charge neutrality is satisfied also in the ash by quarks and electrons as
\begin{eqnarray*}
 \frac{1}{3} (2 n_{up}- n_{dn}- n_{sg} ) =n_{e}.
\end{eqnarray*}

\subsection{Deconfinement \label{model_crid}} 
In this paper we consider the shock-induced conversion scenario, in which shock compression raises the density of HM to the critical density, at which  the deconfinement to 2QM occurs via strong interactions alone. Since the latter process is nothing but a phase transition in equilibrium, the critical density is obtained from the following conditions:
\begin{eqnarray}
P_{H} &=&P_{2QM}  \label{eq:cri1}, \\ 
T_{H} &=&T_{2QM} \label{eq:cri2}, \\  
\mu _{p} &=& 2\mu _{up} + \mu _{dn}  \label{eq:cri3},  \\
\mu _{n} &=& \mu _{up} + 2\mu _{dn}  \label{eq:cri4}.
\end{eqnarray}
The subscripts $2QM$ and $H$ 
indicate the quantities for 2QM and HM, respectively.
We assume that the phase transition is of the first order and hence it occurs via the mixed phase, in which HM and 2QM co-exist.

To describe the mixed phase, we define the volume fraction, $r$, of QM. 
Note that  strange quarks are not present in  the QM, since  deconfinement is accomplished on the time scale of strong interactions. 
 Strange quarks start to populate only on the time scale of weak interactions, on the other hand, since they are produced only via weak interactions.

If the shock wave is weak, the deconfinement process may not be completed 
and  the volume of 2QM  may reach only $r = r_{2QM} < 1$  just after the shock passage.
Then 
strange quarks start to emerge only  in the volume that 2QM occupies 
 and the resultant 3QM occupies the whole volume only later on the time scale of weak interactions. 
If the shock wave is even weaker, a mixed state of  HM and 3QM may result as the final state with a volume fraction of 3QM less than unity: $r = r_{3QM}$ ($r_{2QM}<r_{3QM}<1$).

The basic equations of hydrodynamics are the same as  for the toy model, Eqs.~(\ref{eq:eq4}-\ref{eq:eq6}). 
We start integration from the initial state, employing the EOS for 
 HM up to the critical density, at which deconfinement commences. 
Then we continue integration, taking into account the conditions for the phase equilibrium between HM and 2QM, Eqs.~(\ref{eq:cri1}-\ref{eq:cri4}), as well as 
the conservations of individual quarks expressed  as 
\begin{eqnarray}
Y_{up} &=& (1-r)\frac{2n_p + n_n }{3n_B }+r\frac{n_{up}}{3n_B }= {\rm const.,} 
\end{eqnarray}
 where $Y_{up}$ is the fraction of up quarks.
In this mixed phase, we assume that electrons are uniform ($\mu _{el}^H = \mu _{el}^Q $) 
and that  charge neutrality is valid only globally: $n_{el} = (1-r) n_{p}+r (2 n_{up}-n_{dn})/3$.
Neutrinos essentially do not interact with other particles during   deconfinement and their number density and temperature do not change.
It  hence occurs 
that neutrinos and matter have different temperatures just after the passage of shock wave.

Note that it was often assumed in the literature \cite{lugones98,iida98,bombaci04} that $Y_{up}^H = Y_{up}^Q$ and  $Y_{dn}^H = Y_{dn}^Q$ holds in the mixed states instead of Eq.~(\ref{eq:cri3}) and (\ref{eq:cri4}) employed in this paper. The authors of these papers justified their assumption, stating that strong interactions conserve quark flavors. The last statement is certainly true but that does not necessarily mean that the fractions of up and down quarks are identical between the HM and 2QM that coexist. As a matter of fact, our condition, Eq.~(\ref{eq:cri3}) and (\ref{eq:cri4}), leads in general to different fractions in the two coexisting phases. This is possible without changing quark flavors if neutrons are deconfined more easily than protons or vice versa. Since neutrons have larger chemical potentials in our models, the former is true in this paper. It should be also pointed out that the mixed states with a different flavor abundance in each coexisting phase have lower free energies than those with the identical fractions of flavors as long as the surface and Coulomb energies are ignored. Without reliable estimates of these energies we cannot say which condition is more suitable. We hence adopt Eq.~(\ref{eq:cri3}) and (\ref{eq:cri4}), which are consistent with the rest of the paper, as a criterion for the deconfinement of HM to 2QM in this paper.

\subsection{$\beta$-equilibration in 3QM \label{modeleq}}
After the complete ($r=1$) or incomplete $(r=r_{2QM} < 1)$ deconfinement,
  strange quarks start to populate in  QM on the time scale of $\beta$-equilibration, $\tau$.
The neutrino temperature, $T_{lep}$, and lepton fractions are assumed to change gradually toward the equilibrium values
on the same time scale, which are approximately described in this paper as follows:
\begin{eqnarray}
u\frac{dY_{lep}}{dx} &=& \frac{Y_{lep}^{eq}-Y_{lep}}{\tau}  \label{eq:wk1}, \\
u\frac{dT_{lep}}{dx} &=& \frac{T_{lep}^{eq}-T_{lep}}{\tau} \label{eq:wk2}, \\
u\frac{dY_{up}}{dx} &=& \frac{Y_{up}^{eq}-Y_{up}}{\tau} \label{eq:wk3}, \\
u\frac{df_s}{dx} &=& \frac{f_s^{eq}-f_s}{\tau} \label{eq:wk4}.
\end{eqnarray}

In the incomplete-deconfinement case, strange quarks appear only in the volume occupied by QM,
 as noted in the previous subsection. 
As the fraction of strange quarks rises toward  the $\beta$-equilibrium, the volume fraction of QM, $r$, also increases either from $r=r_{2QM}$ to $r=1$ or to $r=r_{3QM}$  on the time scale of weak interactions.
We solve  Eqs.~(\ref{eq:wk1}-\ref{eq:wk4}) together with the equilibrium conditions between HM and 3QM, which are given by $P_{H} =P_{3QM}$, $T_{H} =T_{3QM}$ and Eqs.~(\ref{eq:cri3}-\ref{eq:cri4}).

\section{Results of Realistic Models \label{result}}
In the following we present the numerical results obtained for the realistic models with the parameters listed in Table~\ref{table1} and also shown in Fig.~\ref{alphabag}. Conversions from PNS matter ($T=10$MeV, $Y_{lep}$=0.3) are first discussed and those from cold NS matter are considered thereafter. Examples of both the complete- and incomplete deconfinements are given for these cases. In all the models the normalized viscosity is set to $\bar{\nu}=1.0$, which is larger than the realistic value by many orders of magnitude, so that the shock width could be widened and well resolved numerically. We give some comments on the results with smaller (but still tractable) viscosities and their implications at the end of this section.

\begin{figure}[t]
\includegraphics[width=12cm]{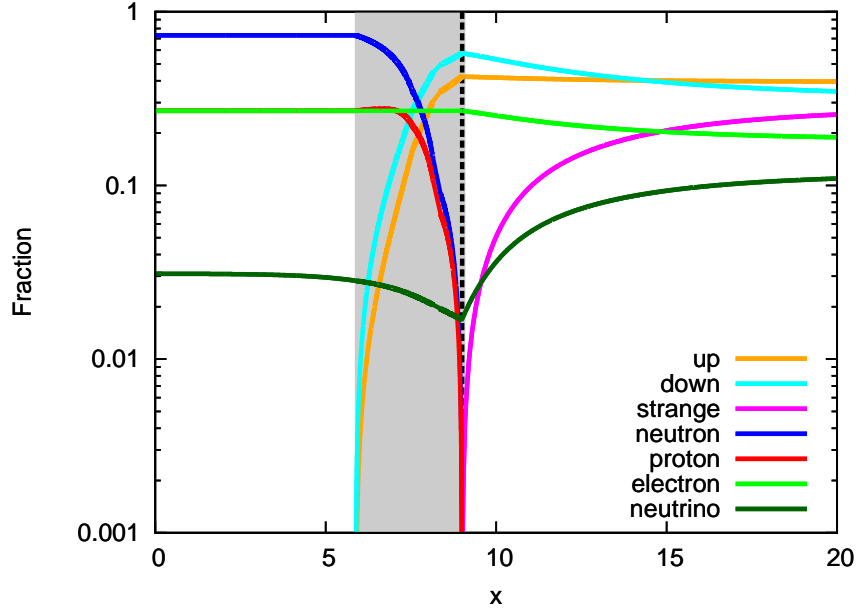}
\caption{Fractions of various particles for the model with $B^{1/4}=140$~MeV and  $\alpha_s=0.40$.
Each fraction is defined to be the ratio of the number density of the particle to the baryonic number density. The $x$ coordinate is normalized by the typical length of weak interactions,
$\tau \times c_s$. The shaded region stands for the deconfinement region.
The dashed line indicates the end of shock wave. }
\label{frac_sm}
\end{figure}
\begin{figure}[h!]
\includegraphics[width=12cm]{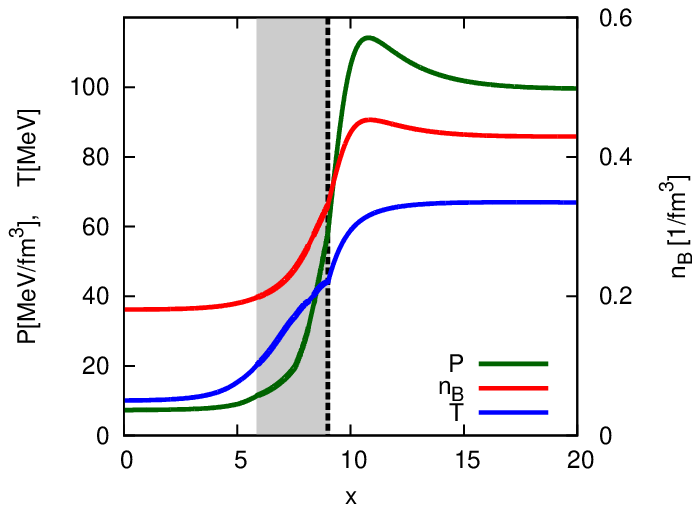}
\caption{The pressure, baryonic number density and temperature for the model with $B^{1/4}=140$~MeV and  $\alpha_s=0.40$. The shaded regions stand for the deconfinement region.
The dashed line indicates the end of shock wave. }
\label{pnbt_sm}
\end{figure}

We begin with the model with $B^{1/4}=140$ MeV and $\alpha_s = 0.4$, which is an example that satisfies the requirement for the SQM hypothesis. The initial HM is a PNS matter with the density of $\rho= 3.0 \times 10^{14}$~$\rm{g/cm}^3$ 
and the Mach number of  $M_i=3.0$.
Fig.\ref{frac_sm} shows the fractions of various particles, which are defined as $n_i/n_B$ with $n_i$ being the number density of particle $i$.
Once the critical density is reached at 
$x = 5.9$.
 2QM grows very rapidly  in the sea of the HM  until the latter
disappears completely. This model is hence an example to give the complete deconfinement. 
Although the deconfinement process should occur on the
time scale of strong interactions
and the deconfinement region is extremely thin in reality, it is extended artificially in our models by adopting the large viscosity.
Once
the deconfinement is completed at $x = 9.0$, strange quarks increase quickly at first and
their fraction approaches the asymptotic value rather slowly later.

The evolutions of thermodynamical quantities are shown in Fig.\ref{pnbt_sm}.
The number density and pressure increase by shock compression up to $x \sim 11.0$.
After the completion of deconfinement at $x=9.0$,
the increment of  temperature  changes  due to the appearance of strange quarks.
Then the total pressure
 and number density are reduced after shock passage at $x \gtrsim 11.0$,
 since the Fermi energies of other quarks decrease. 
On the other hand,  the temperature still increases even in this phase. This is due to the entropy production in the irreversible  $\beta$-equilibration.

\begin{figure}[t]
\includegraphics[width=8cm]{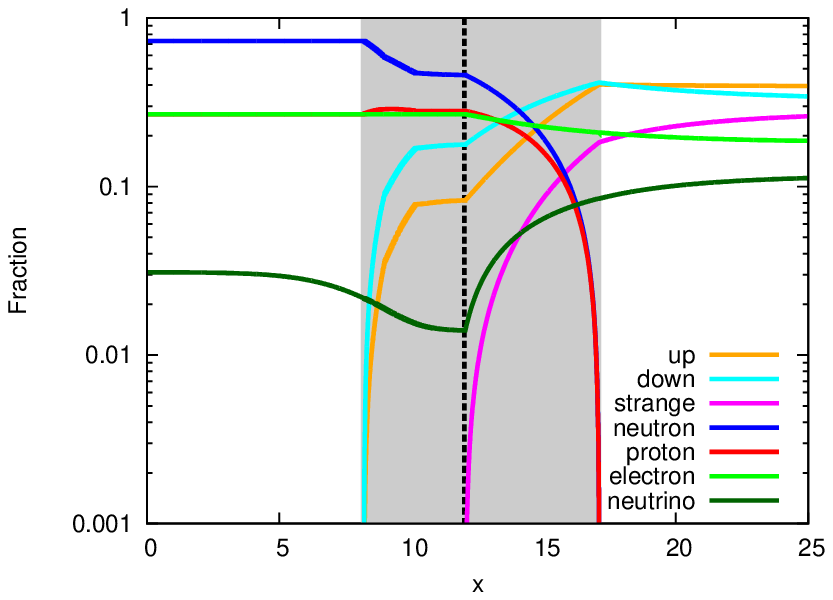}
\includegraphics[width=8cm]{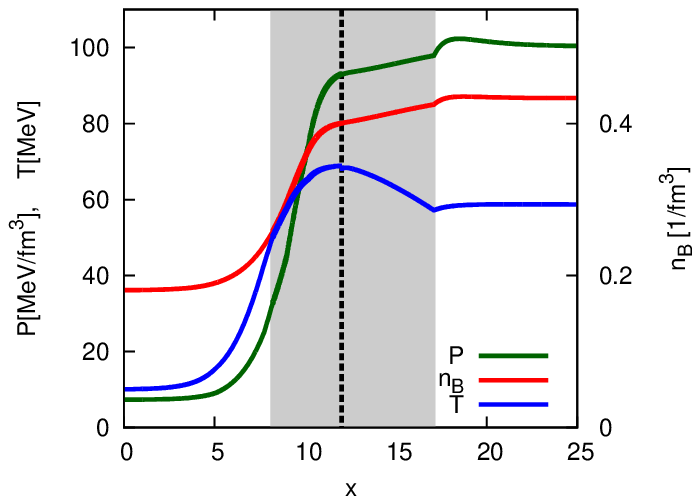}
\caption{Fractions of various particles
(left) and the pressure, baryonic number density and temperature (right) for the model with $B^{1/4}=140$~MeV and  $\alpha_s=0.60$.
The shaded regions and dashed lines  stand for the deconfinement region
and end of shock wave, respectively.
The incomplete deconfinement occurs in this case.
}
\label{typeb}
\includegraphics[width=8cm]{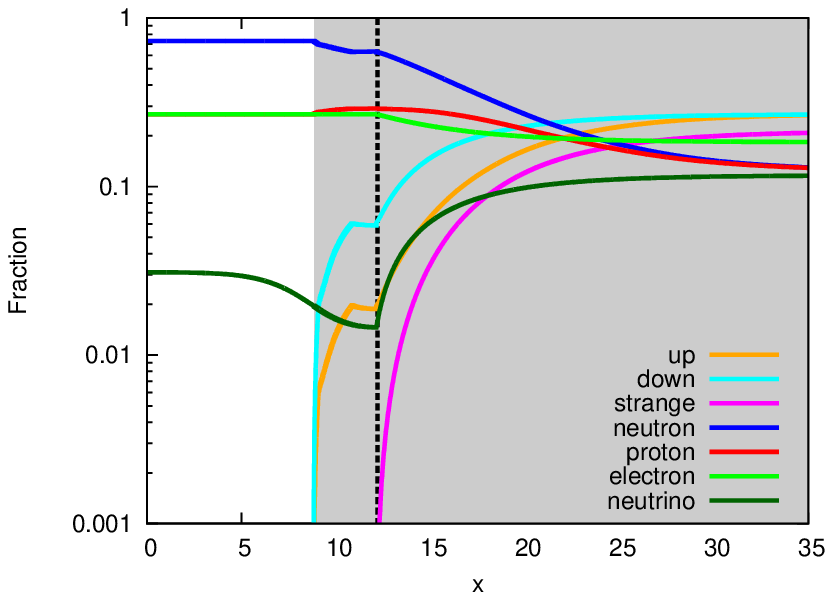}
\includegraphics[width=8cm]{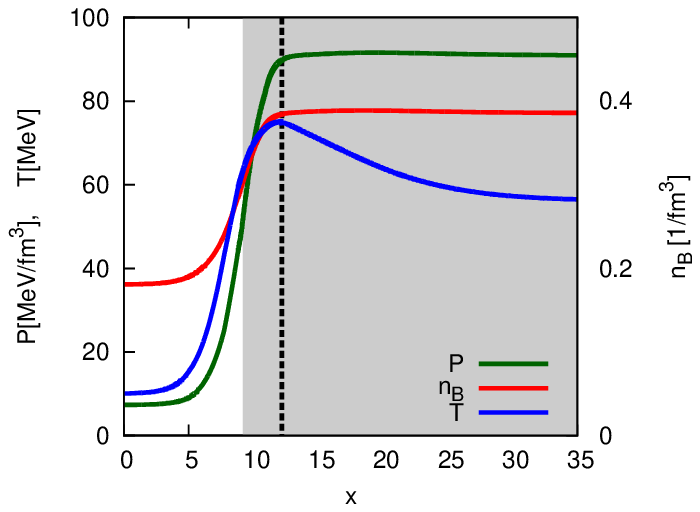}
\caption{The same as Fig.~\ref{typeb} but for the model  with $B^{1/4}=130$~MeV and  $\alpha_s=0.80$.
The incomplete deconfinement occurs and the final state is in the mixed phase of HM and 3QM in this case.}
\label{typec}
\end{figure}

The deconfinement process is not completed in the shock wave for
 the model with $B^{1/4}=140$ MeV and $\alpha_s$=0.60, which is displayed  in Fig.~\ref{typeb}.
This happens because the harder 
 2QM  prevents the shock wave from 
being compressed sufficiently
to complete deconfinement.
Just after the shock passage ($x\sim12.0$), the fraction of the volume that 2QM occupies is $r_{2QM}=0.29$. 
The conversion from 2QM to 3QM by the generation of strange quarks occurs only  in this volume 
surrounded by HM. 
The volume fraction of QM increases gradually then
on the time scale of weak interactions as the strangeness is accumulated just like the diffusion-induced conversion, which will be studied in detail in the forthcoming paper, 
 and 
QM occupies the entire volume, i.e., $r=1$ is reached 
at $x=17.3$. 
The density increase 
after the appearance of strange quarks is slower in this incomplete-deconfinement case 
 than in the complete-deconfinement  model with $B^{1/4}=140$ MeV and $\alpha_s$=0.40. 
This is  due to the coexistence of QM with HM  at $12.0 < x < 17.3$.
The  width of conversion region (see  Fig.~\ref{fig_structure}),
 $\lambda_w \sim 13.0$ $(12.0 < x \lesssim 25.0)$ is accordingly larger in this model than  that for the complete-deconfinement model, $\lambda_w \sim 11.0$  $(9.0 < x \lesssim 20.0)$.
The temperature is 
lowered as the volume fraction of QM increases at $12.0<x<17.3$
due to negative latent energies associated with the conversion from HM to 3QM.

The incomplete deconfinement observed above 
also occurs
for $B^{1/4}=140$MeV and $\alpha_s=0.4$ 
  if the initial Mach number is smaller.
The model with $B^{1/4}=130$ MeV and $\alpha_s$=0.80 also gives the incomplete deconfinement as shown in Fig.~\ref{typec}.
Just after the shock passage, the volume fraction of 2QM is 
$r_{2QM}=0.11$ at $x = 12.2$.
It turns out that  in this model, 
 3QM does not occupy 
 the whole volume even in the final state and the terminal volume fraction of 3QM is 
 $r_{3QM}=0.78$.
This happens because 3QM is substantially harder in this model than in the previous cases.
The width of the conversion region, $\lambda_w \sim 20.0$, is also considerably larger than those of the other two models, $\lambda_w \sim$ 11.0 and 13.0.

\begin{figure}[t]
\includegraphics[width=12cm]{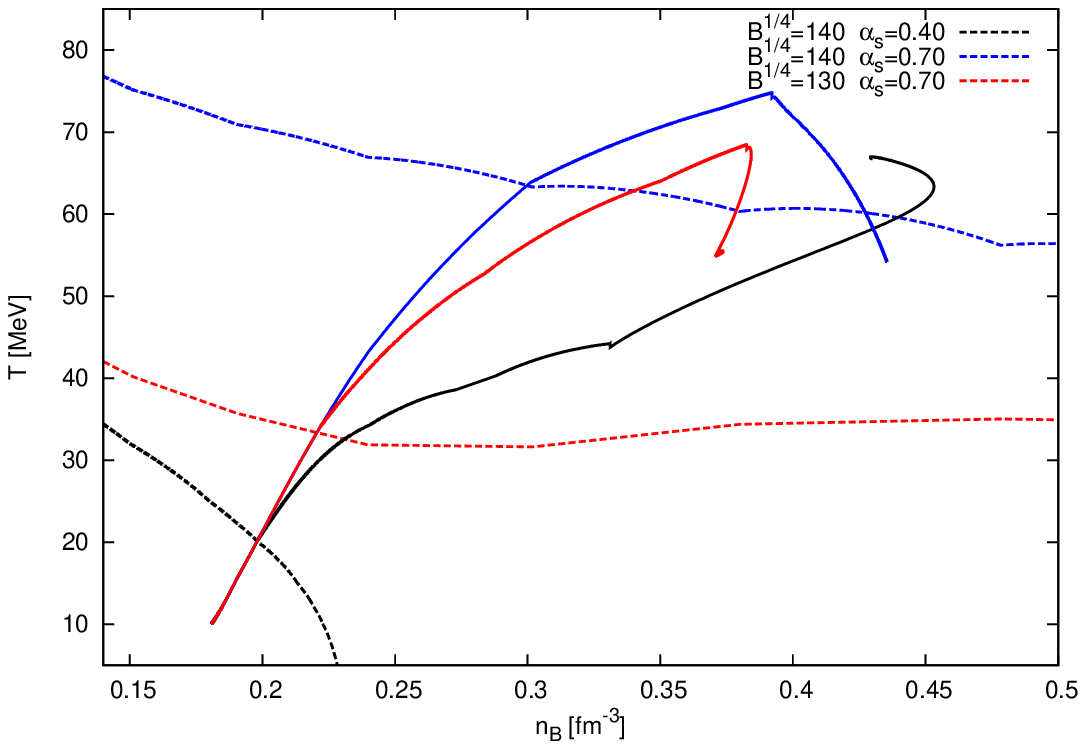}
\caption{The trajectories in the $n_B$-$T$ plane for 
 the models, for which  $B^{1/4}$ and  $\alpha_s$ are 130~MeV and 0.70 (red), 140~MeV and 0.70 (blue), 140~MeV and 0.40 (black). Dashed lines indicate the critical densities,
 at which deconfinement from HM to 2QM occurs for these EOS parameters.
  }
\label{compeos}
\end{figure}

\begin{figure}[t]
\includegraphics[width=12cm]{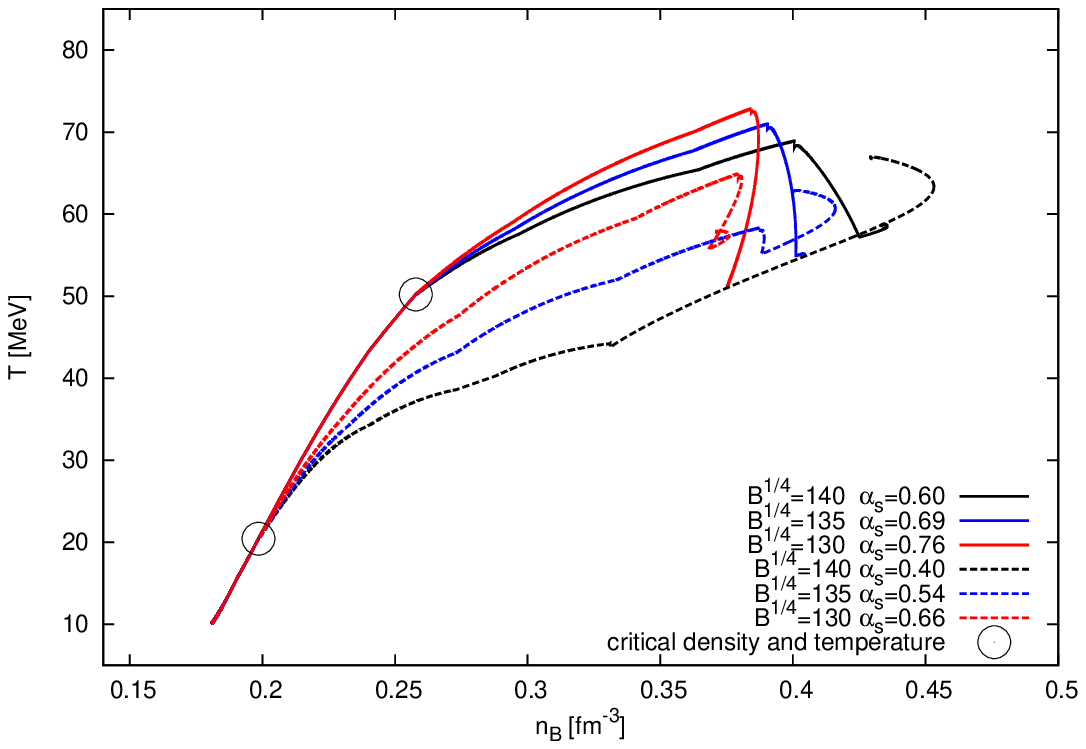}
\caption{The trajectories in the $n_B$-$T$ plane for 
 the models with  $B^{1/4}=$  are 130~MeV (red), 135~MeV (blue) and 140~MeV (black),
 which have the same critical densities: $n_B=0.20$ (dashed lines) or 0.26 (solid lines) fm$^{-3}$, which are marked with circles.}
\label{compeos2}
\end{figure}

We now discuss the systematics in detail.
The critical density is higher for 
 models with larger $B^{1/4} $ and/or $\alpha_s$ as already explained in \ref{modeleos}, which can also be confirmed by the dashed lines in Fig.~\ref{compeos}.
On the other hand, $B^{1/4}$ and $\alpha_s$ affect the stiffness of EOS differently.
The pressure for a fixed number density  decreases with $B^{1/4}$ 
whereas it increases with  $\alpha_s$. 

 Figure~\ref{compeos} compares the results of the three models with   $B^{1/4}=$ 140~MeV and  $\alpha_s=$~0.70,
$B^{1/4}=$ 130  and  $\alpha_s=$~0.70, 
and $B^{1/4}=$ 140~MeV and  $\alpha_s=$~0.40.
The former two have incomplete deconfinement whereas the last one has complete deconfinement as has been already presented.
The final state  of  $B^{1/4}=$ 140~MeV  and  $\alpha_s=$~0.70 is  a mixed phase of  HM and 3QM 
just as the model with $B^{1/4}=130$ MeV and $\alpha_s=$~0.80 shown in 
 Fig.~\ref{typec},
whereas pure 3QM results for 
 the model with $B^{1/4}=$130~MeV  and  $\alpha_s=$~0.70  as for the model in Fig.~\ref{typeb}.
It is found that the final densities are higher for the models with larger $B^{1/4}$ because the EOS is softer then.
The final temperatures, on the other hand,  are 
lower for  larger $\alpha_s$ because the (absolute value of the negative) latent 
energy for deconfinement is greater.
These features do not depend on the critical density as shown in  Fig.~\ref{compeos2}, where we compare  different models 
that have the same critical densities $n_c=$ 0.2 or 0.26 fm$^{-3}$.
We can confirm  that the final density depends only on $B^{1/4}$ and the final temperatures is lower for larger  $\alpha_s$.
Whether deconfinement is completed or not in the shock wave
also depends on $B^{1/4}$ and $\alpha_s$.
 In fact, if we choose the pairs so that they would give the same critical density, how deconfinement is terminated still depends on the bag constant.
Among  three models with $n_c=$ 0.2 fm$^{-3}$ in Fig.~\ref{compeos2}, the model with $B^{1/4}=140$~MeV results in the complete deconfinement 
whereas the other models lead to  the incomplete deconfinement.
The width of  conversion  region also depends on the stiffness of EOS.
For the instance, the model with $B^{1/4}$=130 MeV  has 
about 1.2 times wider a conversion region
than the model with $B^{1/4}$=140 MeV does for the case of  $n_c=$ 0.26 fm$^{-3}$. 

\begin{figure}[t]
\includegraphics[width=10cm]{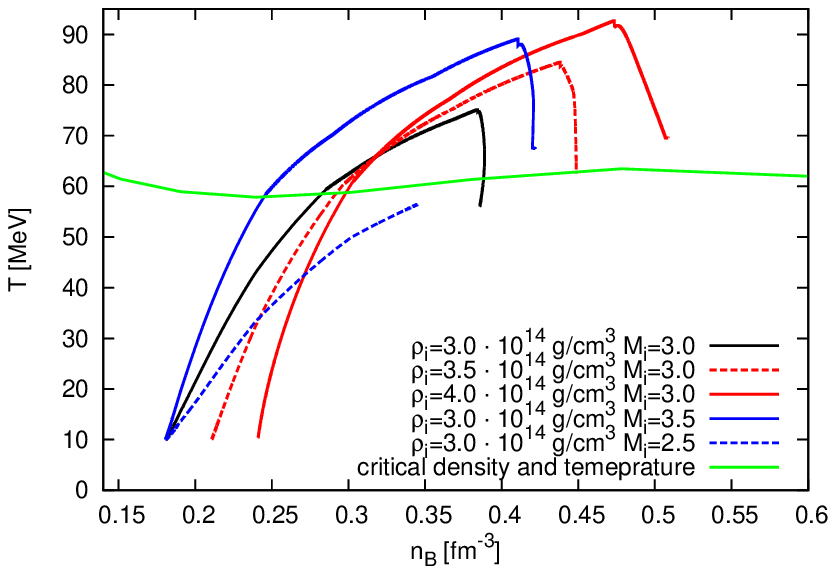}
\caption{
The same as Figs.~\ref{compeos} and~\ref{compeos2} but for 
 the models with different initial Mach numbers, $M_i$, and/or densities, $\rho_i$. 
The bag and strong coupling constants are 
$B^{1/4}=$ 130~MeV and  $\alpha_s=$ 0.80, respectively.
The black solid line shows the reference model with  $\rho_i= 3.0 $ g/cm$^3$ and $M_i=3.0$.
The models with $\rho_i= $3.5 and 4.0 g/cm$^3$ are shown in red
whereas those  with  $M_i=$ 2.5 and 2.0 are displayed in blue.
The critical densities for this combination of the EOS parameters are indicated with the green solid line.
}
\label{compvelo}
\end{figure}

\begin{figure}[t]
\includegraphics[width=10cm]{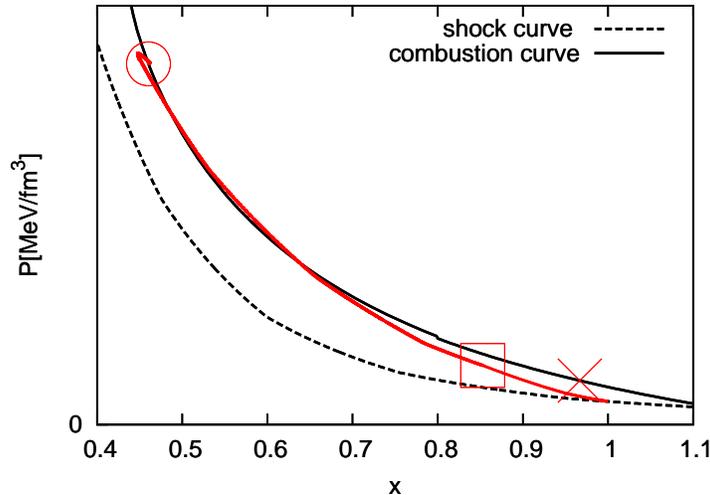}
\caption{
The Hugoniot curves for combustion (solid line) and shock
wave (dashed line) in the realistic model  with $B^{1/4}=$ 125~MeV, $\alpha_s=$ 0.80 and $M_i=3.5$.
The actual trajectory is shown with the red curve.
The final state corresponding to strong (weak) detonation is marked by circle (cross).
The critical point with $\rho=\rho_c$ is also indicated by square.
}
\label{exoreal}
\end{figure}

\begin{figure}[t]
\includegraphics[width=8cm]{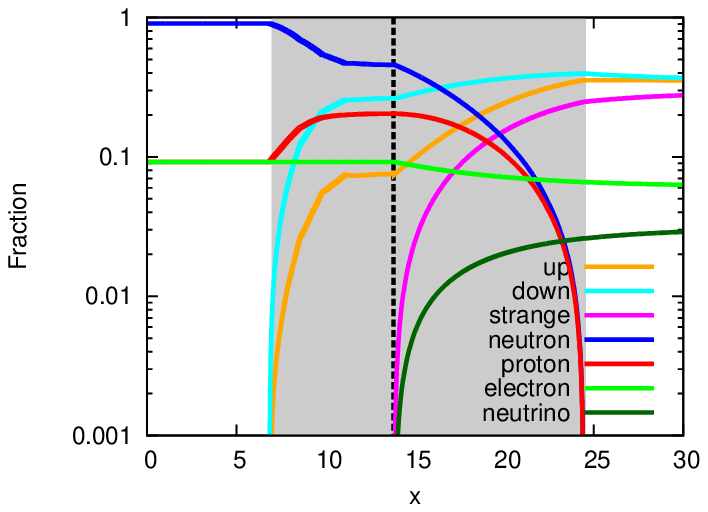}
\includegraphics[width=8cm]{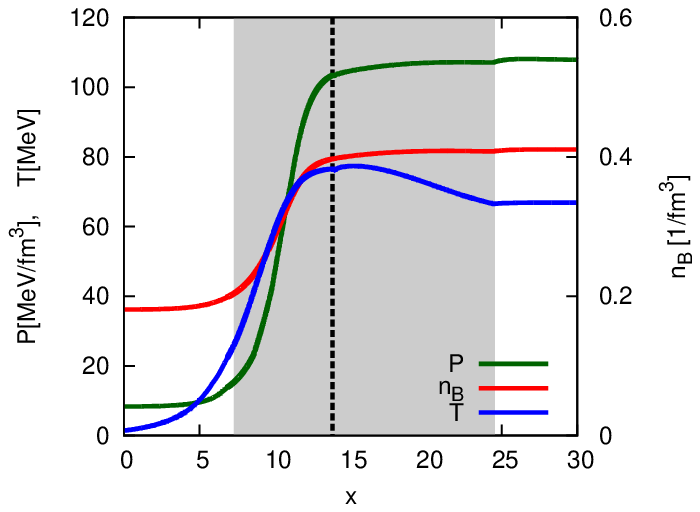}
\includegraphics[width=8cm]{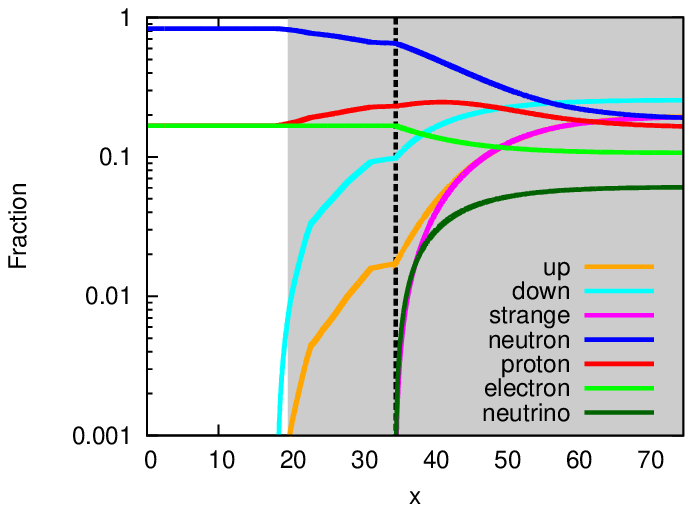}
\includegraphics[width=8cm]{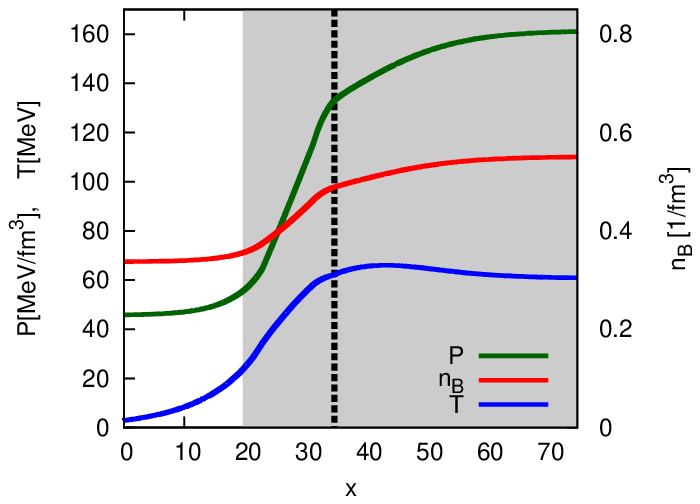}
\caption{
The same as Fig.~\ref{typeb} but for the conversion from the NS matter with $\rho_i= 3.0 $ g/cm$^3$ (top) and 5.6 g/cm$^3$ (bottom). }
\label{nufree}
\end{figure}

Next we pay attention to 
the dependence on the initial Mach number, $M_i$, and density, $\rho_i$, using the model with $B^{1/4}=$ 130 MeV and  $\alpha_s$=0.8, which is shown in Fig.~\ref{compvelo}.
We find that the final state is highly sensitive to 
the initial condition.
For example, the model with $M_i=2.5$ does not reach the critical density and pure HM remains, since the shock compression is weak.
On the other hand, the model  with $M_i=3.5$ obtains  pure 3QM as the final state thanks to  the sufficient compression,
 whereas the asymptotic  state for the  intermediate  model with  $M_i=3.0$ is  found to be in the mixed phase of 3QM and HM as shown in Fig.~\ref{typec}.  
The initial density also affects whether 3QM occupies the entire volume in the end or not.
The models with $\rho_i$ = 3.0 and 3.5  $\times 10^{14}$~g/cm$^3$ 
end up with the mixed states of HM and 3QM and
the final volume fractions are $r_{3QM}$=0.78 and 0.84, respectively. 
The model with an even higher initial density of $\rho_i$ = 4.0$\times 10^{14}$~g/cm$^3$  leads to  pure 3QM or  $r_{3QM} =1.0$.

 We have so far demonstrated that storing detonation always obtains in the realistic models with  $B^{1/4} \gtrsim 130$, which are all in the endothermic regime as shown in Fig.~\ref{fig_hugoniot1}.
This is almost trivial,
 since  matter is decompressed in the weak detonation in the endothermic regime whereas it is compressed in the shock wave.
Fig.~\ref{exoreal} displays the result of the model with  $B^{1/4}=125$~MeV, $\alpha_s=0.80$ and  $M_i=3.5$  as an example in the exothermic regime.
This model has incomplete deconfinement and leads to pure 3QM as  the final state. 
This confirms the conclusion obtained with the toy model that strong detonation is the unique outcome of the shock-induced conversion.

Finally, we mention the results for the conversion 
from  the NS matter, in which $\beta$-equilibrium is established without 
neutrinos in the initial state with $T=0$~MeV. 
 The model parameters are set to be  $B^{1/4}=130$~MeV, $\alpha_s=0.80$, $M_i=2.0$, and $\rho_i =$ 3.0 or  5.6 $\times 10^{14}$~g/cm$^3$.
The results are not much changed from those of the corresponding PNS cases as also shown in Fig.~\ref{nufree}.
Neutrinos, which are absent initially,  start to appear with strange quarks via weak interactions at $x \sim 14$ and $34$ for $\rho_i =$ 3.0 and  5.6 $\times 10^{14}$~g/cm$^3$, respectively,
at which deconfinement is terminated incompletely.
Note that the critical density for the NS matter is smaller in general  than for the PNS matter due to the larger Fermi energies of neutrons.
Deconfinement  hence occurs even in the models with  $\rho_i =$ 3.0  and  $M_i=2.0$ for the NS matter
 although the corresponding PNS matter can not reach the critical density even with $M_i=2.5$ as shown in Fig.~\ref{compvelo}.
In the case of  $\rho_i =$  5.6 $\times 10^{14}$~g/cm$^3$ shock compression is weaker due to greater repulsive forces of HM at such high densities.
The resultant lower temperature  results in a mixed state 
 of  3QM and HM in the final state.

In the calculations done so far, the viscosity is   normalized to unity, which is much
larger than the realistic 
value
as previously noted.
This was necessary to widen the shock width and treat the processes that occur inside the shock wave on the same footing as other events that happen afterward on the time scale of weak interactions.
We have confirmed that smaller but still tractable 
 viscosities 
do not change the results qualitatively although the width of the conversion region tends to get wider in general quantitatively. 
We hence expect that the results obtained in this paper will be still valid for the realistic shock widths.

\section{Conclusion and Discussions \label{conclusion}}
Based on the hydrodynamical description, we have studied the shock-induced conversion of hadronic matter (HM) to three-flavor quark matter (3QM), taking into account the structure inside the shock wave.
In this scenario, the shock wave compresses HM beyond the critical density,
 at which the deconfinement to two-flavor quark matter (2QM) occurs as the first order phase transition in thermal and chemical equilibria on the time scale of strong interactions; then strange quarks are produced via weak interactions and  $\beta$-equilibration ensues. This is an irreversible process accompanied by entropy generation. This series of events together with the matter motion have been described consistently by the hydrodynamical model. Introducing finite viscosities, we have taken into account the structure of the shock wave instead of treating it as a discontinuity as usual. We have employed both the simple toy model and the realistic model, in which microphysics such as EOS is more elaborated to understand the conversion processes.

In the analysis with the toy model, which is meant to elucidate the generic properties of the conversion qualitatively, we have demonstrated, varying model parameters rather arbitrarily in a wide range, that strong detonation almost always obtains both in the exothermic and endothermic regimes, the latter of which has no counter part in terrestrial combustion but is rather common in the conversion of HM to QM.
In our realistic model, we have adopted the EOS based on relativistic mean filed theory for the hot proto-neutron star (PNS) matter as well as  for the cold neutron star (NS) matter and employed the MIT bag model with the first-order perturbation corrections for the EOS of QM. We have confirmed the basic scenario: the shock compression raises the density, pressure and temperature and induces deconfinement at the critical density; once it commences, the mixed state of HM and 2QM is formed with the volume of latter increasing as the density rises; the negative latent energy in deconfinement suppresses the rise in temperature particularly for large $\alpha_s$; after shock passage, strange quarks start to populate via weak interactions and eventually 3QM in  $\beta$-equilibrium is realized.

We have also found that if the shock is not strong enough, deconfinement is not completed inside the shock wave, i.e., the mixed state of HM and 2QM is left behind just after the shock passage. 
Then strange quarks are generated only in the region that 2QM occupies. 
As the strangeness increases in that region, the volume of the region itself  also increases, since such state is more stable thermodynamically.
For the pairs of bag constant and strong coupling constant that satisfy the constraints that the critical density should be larger than the nuclear saturation density and the maximum mass of cold quark star should be larger than 2M$_\odot$, this incomplete deconfinement seems more likely than otherwise. 
It is also interesting that the final state may also be a mixed state of HM and 3QM, depending on the initial density and Mach number. 
We have found no essential difference between the conversions from the hot PNS matter and those from the cold NS matter except that
 the critical density tends to be lower for the latter.

We have seen that the shock-induced conversion is more likely to occur in the endothermic regime if one considers various constraints (see Fig.~\ref{alphabag}) and argued that there is no a priori reason to discard the conversion in that regime. If this is really the case, strong detonation will be the unique outcome of the shock-induced conversion, since matter is compressed in the shock wave whereas it would be decompressed in the weak detonation for the endothermic regime. We have observed, however, that strong detonation obtains also in the exothermic regime in our realistic model, the conclusion also supported by the phase space analysis for the toy model. It is interesting to point out that there is no Jouget point in the endothermic combustion. It is well known that the so-called Zeldovich detonation, which is a self-similar combustion that occurs at the Jouget point in the detonation branch, obtains for the single-point-ignition in a uniform fuel. It is an intriguing question then what happens to such a self-similar propagation of detonation wave in case there is no Jouget point, which may have some implications for the global conversion of NS or PNS by detonation.

In this paper, we have employed 
 the non-relativistic formulation for simplicity. 
This is certainly not a good approximation in the current context,
since  the conversion front travels at a fraction of $c$.
It is still not so high as to affect the results qualitatively, though. In fact, it
is straightforward to accommodate special relativity as already demonstrated in section \ref{jump}
 except for the subtle problem with causality in viscous hydrodynamics and some preliminary studies done so far indicate that the main results in this paper are not changed qualitatively indeed. 
The full results  will be presented in the sequel to this paper.

It is repeated that our analysis is local, with the front velocity being regarded as a free parameter. In reality, it is determined uniquely by the global configuration of neutron star and appropriate boundary conditions. Our standpoint in this local analysis is that we should list up all possible inner structures of the conversion region although some of them may not be realized in reality. Put another way, whatever conversion front appears in global simulation should be included in our list if it is observed locally. Such an approach is complementary to the global analysis by large-scale simulations. We believe that both of them are indispensable to obtain a coherent picture of the conversion from neutron stars to quark stars.
Muons  have been neglected   as a leptonic component in this paper, which may not be justified completely, particularly for  the conversion from  the NS matter.
We believe, however, that the  results in this paper will not be changed
 qualitatively by the inclusion of muons, since they are not so abundant
at rather low densities considered above:
their abundances are estimated to be  $\sim$ 2.7 and  $7.8 \%$ at $\rho =$ 3.0 and 5.6 $\times 10^{14}$~g/cm$^3$,
respectively. 

There are many uncertainties in the EOS of QM other than the bag and strong coupling constants: at present there supposed to be a rich variety of phases 
 such as color-flavor locking \cite{lugones2002,lugones2003,horvath2004,kovacs2009} and 
quark clustering in cold quark matter  \cite{lai2009,lai2010}.
We have also to take into consideration the surface effect in dealing with the mixed phase of HM and QM, since it may affect the dynamics in that phase.
Last but not least, other conversion scenarios should be investigated. As mentioned in section \ref{cartoon_energy}, 
 the diffusion-induced conversion
has been discussed in the literature over the years. We are currently investigating it with the same approach.
The results of the detailed analysis will be reported in our next paper.

\begin{acknowledgments}
The authors thank T. Hasegawa  for the useful discussion. 
A part of the numerical calculations were carried out on  PC cluster at Center
for Computational Astrophysics, National Astronomical Observatory of Japan.
 This work was partially supported by Grant-in-Aid for the Scientific Research
from the Ministry of Education, Culture, Sports, Science
and Technology (MEXT), Japan (24103006, 24244036).
\end{acknowledgments}

\bibliography{reference}

\begin{thebibliography}{63}
\expandafter\ifx\csname natexlab\endcsname\relax\def\natexlab#1{#1}\fi
\expandafter\ifx\csname bibnamefont\endcsname\relax
  \def\bibnamefont#1{#1}\fi
\expandafter\ifx\csname bibfnamefont\endcsname\relax
  \def\bibfnamefont#1{#1}\fi
\expandafter\ifx\csname citenamefont\endcsname\relax
  \def\citenamefont#1{#1}\fi
\expandafter\ifx\csname url\endcsname\relax
  \def\url#1{\texttt{#1}}\fi
\expandafter\ifx\csname urlprefix\endcsname\relax\def\urlprefix{URL }\fi
\providecommand{\bibinfo}[2]{#2}
\providecommand{\eprint}[2][]{\url{#2}}

\bibitem[{\citenamefont{{Baym}}(2007)}]{baym2006}
\bibinfo{author}{\bibfnamefont{G.}~\bibnamefont{{Baym}}}, in
  \emph{\bibinfo{booktitle}{Quark Confinement and the Hadron Spectrum VII}},
  edited by \bibinfo{editor}{\bibnamefont{{J.~E.~F.~T.~Ribeiro, N.~Brambilla,
  A.~Vairo, K.~Maung, \& G.~M.~Prosperi }}} (\bibinfo{year}{2007}), vol.
  \bibinfo{volume}{892} of \emph{\bibinfo{series}{American Institute of Physics
  Conference Series}}, pp. \bibinfo{pages}{8--14},
  \eprint{arXiv:nucl-th/0612021}.

\bibitem[{\citenamefont{{Itoh}}(1970)}]{itoh70}
\bibinfo{author}{\bibfnamefont{N.}~\bibnamefont{{Itoh}}},
  \bibinfo{journal}{Progress of Theoretical Physics}
  \textbf{\bibinfo{volume}{44}}, \bibinfo{pages}{291} (\bibinfo{year}{1970}).

\bibitem[{\citenamefont{{Alcock} et~al.}(1986)\citenamefont{{Alcock}, {Farhi},
  and {Olinto}}}]{alcock}
\bibinfo{author}{\bibfnamefont{C.}~\bibnamefont{{Alcock}}},
  \bibinfo{author}{\bibfnamefont{E.}~\bibnamefont{{Farhi}}}, \bibnamefont{and}
  \bibinfo{author}{\bibfnamefont{A.}~\bibnamefont{{Olinto}}},
  \bibinfo{journal}{\apj} \textbf{\bibinfo{volume}{310}}, \bibinfo{pages}{261}
  (\bibinfo{year}{1986}).

\bibitem[{\citenamefont{{Witten}}(1984)}]{witten}
\bibinfo{author}{\bibfnamefont{E.}~\bibnamefont{{Witten}}},
  \bibinfo{journal}{\prd} \textbf{\bibinfo{volume}{30}}, \bibinfo{pages}{272}
  (\bibinfo{year}{1984}).

\bibitem[{\citenamefont{{Madsen}}(1988)}]{madsen}
\bibinfo{author}{\bibfnamefont{J.}~\bibnamefont{{Madsen}}},
  \bibinfo{journal}{Physical Review Letters} \textbf{\bibinfo{volume}{61}},
  \bibinfo{pages}{2909} (\bibinfo{year}{1988}).

\bibitem[{\citenamefont{{Caldwell} and {Friedman}}(1991)}]{friedman}
\bibinfo{author}{\bibfnamefont{R.~R.} \bibnamefont{{Caldwell}}}
  \bibnamefont{and} \bibinfo{author}{\bibfnamefont{J.~L.}
  \bibnamefont{{Friedman}}}, \bibinfo{journal}{Physics Letters B}
  \textbf{\bibinfo{volume}{264}}, \bibinfo{pages}{143} (\bibinfo{year}{1991}).

\bibitem[{\citenamefont{{Vucetich} and {Horvath}}(1998)}]{vucetich}
\bibinfo{author}{\bibfnamefont{H.}~\bibnamefont{{Vucetich}}} \bibnamefont{and}
  \bibinfo{author}{\bibfnamefont{J.~E.} \bibnamefont{{Horvath}}},
  \bibinfo{journal}{\prd} \textbf{\bibinfo{volume}{57}}, \bibinfo{pages}{5959}
  (\bibinfo{year}{1998}), \eprint{arXiv:astro-ph/9802363}.

\bibitem[{\citenamefont{{de Rujula} and {Glashow}}(1984)}]{rujula}
\bibinfo{author}{\bibfnamefont{A.}~\bibnamefont{{de Rujula}}} \bibnamefont{and}
  \bibinfo{author}{\bibfnamefont{S.~L.} \bibnamefont{{Glashow}}},
  \bibinfo{journal}{\nat} \textbf{\bibinfo{volume}{312}}, \bibinfo{pages}{734}
  (\bibinfo{year}{1984}).

\bibitem[{\citenamefont{{Cecchini} et~al.}(2008)\citenamefont{{Cecchini},
  {Cozzi}, {di Ferdinando}, {Errico}, {Fabbri}, {Giacomelli}, {Giacomelli},
  {Giorgini}, {Kumar}, {McDonald} et~al.}}]{cecchini}
\bibinfo{author}{\bibfnamefont{S.}~\bibnamefont{{Cecchini}}},
  \bibinfo{author}{\bibfnamefont{M.}~\bibnamefont{{Cozzi}}},
  \bibinfo{author}{\bibfnamefont{D.}~\bibnamefont{{di Ferdinando}}},
  \bibinfo{author}{\bibfnamefont{M.}~\bibnamefont{{Errico}}},
  \bibinfo{author}{\bibfnamefont{F.}~\bibnamefont{{Fabbri}}},
  \bibinfo{author}{\bibfnamefont{G.}~\bibnamefont{{Giacomelli}}},
  \bibinfo{author}{\bibfnamefont{R.}~\bibnamefont{{Giacomelli}}},
  \bibinfo{author}{\bibfnamefont{M.}~\bibnamefont{{Giorgini}}},
  \bibinfo{author}{\bibfnamefont{A.}~\bibnamefont{{Kumar}}},
  \bibinfo{author}{\bibfnamefont{J.}~\bibnamefont{{McDonald}}},
  \bibnamefont{et~al.}, \bibinfo{journal}{European Physical Journal C}
  \textbf{\bibinfo{volume}{57}}, \bibinfo{pages}{525} (\bibinfo{year}{2008}),
  \eprint{0805.1797}.

\bibitem[{\citenamefont{{Bombaci} et~al.}(2004)\citenamefont{{Bombaci},
  {Parenti}, and {Vida{\~n}a}}}]{bombaci04}
\bibinfo{author}{\bibfnamefont{I.}~\bibnamefont{{Bombaci}}},
  \bibinfo{author}{\bibfnamefont{I.}~\bibnamefont{{Parenti}}},
  \bibnamefont{and}
  \bibinfo{author}{\bibfnamefont{I.}~\bibnamefont{{Vida{\~n}a}}},
  \bibinfo{journal}{\apj} \textbf{\bibinfo{volume}{614}}, \bibinfo{pages}{314}
  (\bibinfo{year}{2004}), \eprint{astro-ph/0402404}.

\bibitem[{\citenamefont{{Mallick} and {Sahu}}(2014)}]{mallick14}
\bibinfo{author}{\bibfnamefont{R.}~\bibnamefont{{Mallick}}} \bibnamefont{and}
  \bibinfo{author}{\bibfnamefont{P.~K.} \bibnamefont{{Sahu}}},
  \bibinfo{journal}{Nuclear Physics A} \textbf{\bibinfo{volume}{921}},
  \bibinfo{pages}{96} (\bibinfo{year}{2014}).

\bibitem[{\citenamefont{{Benvenuto} et~al.}(1989)\citenamefont{{Benvenuto},
  {Horvath}, and {Vucetich}}}]{benvenuto89}
\bibinfo{author}{\bibfnamefont{O.~G.} \bibnamefont{{Benvenuto}}},
  \bibinfo{author}{\bibfnamefont{J.~E.} \bibnamefont{{Horvath}}},
  \bibnamefont{and}
  \bibinfo{author}{\bibfnamefont{H.}~\bibnamefont{{Vucetich}}},
  \bibinfo{journal}{International Journal of Modern Physics A}
  \textbf{\bibinfo{volume}{4}}, \bibinfo{pages}{257} (\bibinfo{year}{1989}).

\bibitem[{\citenamefont{{Benvenuto} and {Horvath}}(1989)}]{benvenuto}
\bibinfo{author}{\bibfnamefont{O.~G.} \bibnamefont{{Benvenuto}}}
  \bibnamefont{and} \bibinfo{author}{\bibfnamefont{J.~E.}
  \bibnamefont{{Horvath}}}, \bibinfo{journal}{Physical Review Letters}
  \textbf{\bibinfo{volume}{63}}, \bibinfo{pages}{716} (\bibinfo{year}{1989}).

\bibitem[{\citenamefont{{Lugones} et~al.}(1994)\citenamefont{{Lugones},
  {Benvenuto}, and {Vucetich}}}]{lugones94}
\bibinfo{author}{\bibfnamefont{G.}~\bibnamefont{{Lugones}}},
  \bibinfo{author}{\bibfnamefont{O.~G.} \bibnamefont{{Benvenuto}}},
  \bibnamefont{and}
  \bibinfo{author}{\bibfnamefont{H.}~\bibnamefont{{Vucetich}}},
  \bibinfo{journal}{\prd} \textbf{\bibinfo{volume}{50}}, \bibinfo{pages}{6100}
  (\bibinfo{year}{1994}).

\bibitem[{\citenamefont{{Gentile} et~al.}(1993)\citenamefont{{Gentile},
  {Aufderheide}, {Mathews}, {Swesty}, and {Fuller}}}]{gentile}
\bibinfo{author}{\bibfnamefont{N.~A.} \bibnamefont{{Gentile}}},
  \bibinfo{author}{\bibfnamefont{M.~B.} \bibnamefont{{Aufderheide}}},
  \bibinfo{author}{\bibfnamefont{G.~J.} \bibnamefont{{Mathews}}},
  \bibinfo{author}{\bibfnamefont{F.~D.} \bibnamefont{{Swesty}}},
  \bibnamefont{and} \bibinfo{author}{\bibfnamefont{G.~M.}
  \bibnamefont{{Fuller}}}, \bibinfo{journal}{\apj}
  \textbf{\bibinfo{volume}{414}}, \bibinfo{pages}{701} (\bibinfo{year}{1993}).

\bibitem[{\citenamefont{{Dai} et~al.}(1995)\citenamefont{{Dai}, {Peng}, and
  {Lu}}}]{dai}
\bibinfo{author}{\bibfnamefont{Z.}~\bibnamefont{{Dai}}},
  \bibinfo{author}{\bibfnamefont{Q.}~\bibnamefont{{Peng}}}, \bibnamefont{and}
  \bibinfo{author}{\bibfnamefont{T.}~\bibnamefont{{Lu}}},
  \bibinfo{journal}{\apj} \textbf{\bibinfo{volume}{440}}, \bibinfo{pages}{815}
  (\bibinfo{year}{1995}).

\bibitem[{\citenamefont{{Sagert} et~al.}(2009)\citenamefont{{Sagert},
  {Fischer}, {Hempel}, {Pagliara}, {Schaffner-Bielich}, {Mezzacappa},
  {Thielemann}, and {Liebend{\"o}rfer}}}]{sagert2009}
\bibinfo{author}{\bibfnamefont{I.}~\bibnamefont{{Sagert}}},
  \bibinfo{author}{\bibfnamefont{T.}~\bibnamefont{{Fischer}}},
  \bibinfo{author}{\bibfnamefont{M.}~\bibnamefont{{Hempel}}},
  \bibinfo{author}{\bibfnamefont{G.}~\bibnamefont{{Pagliara}}},
  \bibinfo{author}{\bibfnamefont{J.}~\bibnamefont{{Schaffner-Bielich}}},
  \bibinfo{author}{\bibfnamefont{A.}~\bibnamefont{{Mezzacappa}}},
  \bibinfo{author}{\bibfnamefont{F.-K.} \bibnamefont{{Thielemann}}},
  \bibnamefont{and}
  \bibinfo{author}{\bibfnamefont{M.}~\bibnamefont{{Liebend{\"o}rfer}}},
  \bibinfo{journal}{Physical Review Letters} \textbf{\bibinfo{volume}{102}},
  \bibinfo{pages}{081101} (\bibinfo{year}{2009}), \eprint{0809.4225}.

\bibitem[{\citenamefont{{Sagert} et~al.}(2010)\citenamefont{{Sagert},
  {Fischer}, {Hempel}, {Pagliara}, {Schaffner-Bielich}, {Thielemann}, and
  {Liebend{\"o}rfer}}}]{sagert2010}
\bibinfo{author}{\bibfnamefont{I.}~\bibnamefont{{Sagert}}},
  \bibinfo{author}{\bibfnamefont{T.}~\bibnamefont{{Fischer}}},
  \bibinfo{author}{\bibfnamefont{M.}~\bibnamefont{{Hempel}}},
  \bibinfo{author}{\bibfnamefont{G.}~\bibnamefont{{Pagliara}}},
  \bibinfo{author}{\bibfnamefont{J.}~\bibnamefont{{Schaffner-Bielich}}},
  \bibinfo{author}{\bibfnamefont{F.-K.} \bibnamefont{{Thielemann}}},
  \bibnamefont{and}
  \bibinfo{author}{\bibfnamefont{M.}~\bibnamefont{{Liebend{\"o}rfer}}},
  \bibinfo{journal}{Journal of Physics G Nuclear Physics}
  \textbf{\bibinfo{volume}{37}}, \bibinfo{eid}{094064} (\bibinfo{year}{2010}),
  \eprint{1003.2320}.

\bibitem[{\citenamefont{{Fischer} et~al.}(2010)\citenamefont{{Fischer},
  {Sagert}, {Hempel}, {Pagliara}, {Schaffner-Bielich}, and
  {Liebend{\"o}rfer}}}]{fischer2010}
\bibinfo{author}{\bibfnamefont{T.}~\bibnamefont{{Fischer}}},
  \bibinfo{author}{\bibfnamefont{I.}~\bibnamefont{{Sagert}}},
  \bibinfo{author}{\bibfnamefont{M.}~\bibnamefont{{Hempel}}},
  \bibinfo{author}{\bibfnamefont{G.}~\bibnamefont{{Pagliara}}},
  \bibinfo{author}{\bibfnamefont{J.}~\bibnamefont{{Schaffner-Bielich}}},
  \bibnamefont{and}
  \bibinfo{author}{\bibfnamefont{M.}~\bibnamefont{{Liebend{\"o}rfer}}},
  \bibinfo{journal}{Classical and Quantum Gravity}
  \textbf{\bibinfo{volume}{27}}, \bibinfo{pages}{114102}
  (\bibinfo{year}{2010}).

\bibitem[{\citenamefont{{Fischer}
  et~al.}(2011{\natexlab{a}})\citenamefont{{Fischer}, {Sagert}, {Pagliara},
  {Hempel}, {Schaffner-Bielich}, {Rauscher}, {Thielemann}, {K{\"a}ppeli},
  {Mart{\'{\i}}nez-Pinedo}, and {Liebend{\"o}rfer}}}]{fischer2011}
\bibinfo{author}{\bibfnamefont{T.}~\bibnamefont{{Fischer}}},
  \bibinfo{author}{\bibfnamefont{I.}~\bibnamefont{{Sagert}}},
  \bibinfo{author}{\bibfnamefont{G.}~\bibnamefont{{Pagliara}}},
  \bibinfo{author}{\bibfnamefont{M.}~\bibnamefont{{Hempel}}},
  \bibinfo{author}{\bibfnamefont{J.}~\bibnamefont{{Schaffner-Bielich}}},
  \bibinfo{author}{\bibfnamefont{T.}~\bibnamefont{{Rauscher}}},
  \bibinfo{author}{\bibfnamefont{F.-K.} \bibnamefont{{Thielemann}}},
  \bibinfo{author}{\bibfnamefont{R.}~\bibnamefont{{K{\"a}ppeli}}},
  \bibinfo{author}{\bibfnamefont{G.}~\bibnamefont{{Mart{\'{\i}}nez-Pinedo}}},
  \bibnamefont{and}
  \bibinfo{author}{\bibfnamefont{M.}~\bibnamefont{{Liebend{\"o}rfer}}},
  \bibinfo{journal}{\apj} \textbf{\bibinfo{volume}{194}}, \bibinfo{pages}{39}
  (\bibinfo{year}{2011}{\natexlab{a}}), \eprint{1011.3409}.

\bibitem[{\citenamefont{{Fischer}
  et~al.}(2011{\natexlab{b}})\citenamefont{{Fischer}, {Blaschke}, {Hempel},
  {Kl{\"a}hn}, {{\L}astowiecki}, {Liebend{\"o}rfer}, {Mart{\'{\i}}nez-Pinedo},
  {Pagliara}, {Sagert}, {Sandin} et~al.}}]{fischer2011b}
\bibinfo{author}{\bibfnamefont{T.}~\bibnamefont{{Fischer}}},
  \bibinfo{author}{\bibfnamefont{D.}~\bibnamefont{{Blaschke}}},
  \bibinfo{author}{\bibfnamefont{M.}~\bibnamefont{{Hempel}}},
  \bibinfo{author}{\bibfnamefont{T.}~\bibnamefont{{Kl{\"a}hn}}},
  \bibinfo{author}{\bibfnamefont{R.}~\bibnamefont{{{\L}astowiecki}}},
  \bibinfo{author}{\bibfnamefont{M.}~\bibnamefont{{Liebend{\"o}rfer}}},
  \bibinfo{author}{\bibfnamefont{G.}~\bibnamefont{{Mart{\'{\i}}nez-Pinedo}}},
  \bibinfo{author}{\bibfnamefont{G.}~\bibnamefont{{Pagliara}}},
  \bibinfo{author}{\bibfnamefont{I.}~\bibnamefont{{Sagert}}},
  \bibinfo{author}{\bibfnamefont{F.}~\bibnamefont{{Sandin}}},
  \bibnamefont{et~al.}, \bibinfo{journal}{ArXiv e-prints}
  (\bibinfo{year}{2011}{\natexlab{b}}), \eprint{1103.3004}.

\bibitem[{\citenamefont{{Dexheimer} et~al.}(2011)\citenamefont{{Dexheimer},
  {Negreiros}, and {Schramm}}}]{dexheimer2011}
\bibinfo{author}{\bibfnamefont{V.}~\bibnamefont{{Dexheimer}}},
  \bibinfo{author}{\bibfnamefont{R.}~\bibnamefont{{Negreiros}}},
  \bibnamefont{and}
  \bibinfo{author}{\bibfnamefont{S.}~\bibnamefont{{Schramm}}},
  \bibinfo{journal}{ArXiv e-prints}  (\bibinfo{year}{2011}),
  \eprint{1108.4479}.

\bibitem[{\citenamefont{{Kang} et~al.}(2010)\citenamefont{{Kang}, {Wang}, and
  {Zheng}}}]{kang2010}
\bibinfo{author}{\bibfnamefont{M.}~\bibnamefont{{Kang}}},
  \bibinfo{author}{\bibfnamefont{X.-D.} \bibnamefont{{Wang}}},
  \bibnamefont{and} \bibinfo{author}{\bibfnamefont{X.-P.}
  \bibnamefont{{Zheng}}}, \bibinfo{journal}{ArXiv e-prints}
  \textbf{\bibinfo{volume}{15}}, \bibinfo{pages}{515} (\bibinfo{year}{2010}),
  \eprint{1003.0263}.

\bibitem[{\citenamefont{{Glendenning}}(1992)}]{glendenning}
\bibinfo{author}{\bibfnamefont{N.~K.} \bibnamefont{{Glendenning}}},
  \bibinfo{journal}{\prd} \textbf{\bibinfo{volume}{46}}, \bibinfo{pages}{1274}
  (\bibinfo{year}{1992}).

\bibitem[{\citenamefont{{Xu}}(2008)}]{yasutake05}
\bibinfo{author}{\bibfnamefont{R.}~\bibnamefont{{Xu}}},
  \bibinfo{journal}{Modern Physics Letters A} \textbf{\bibinfo{volume}{23}},
  \bibinfo{pages}{1629} (\bibinfo{year}{2008}), \eprint{0802.0648}.

\bibitem[{\citenamefont{{Olinto}}(1991)}]{olinto1991}
\bibinfo{author}{\bibfnamefont{A.}~\bibnamefont{{Olinto}}},
  \bibinfo{journal}{Nuclear Physics B Proceedings Supplements}
  \textbf{\bibinfo{volume}{24}}, \bibinfo{pages}{103} (\bibinfo{year}{1991}).

\bibitem[{\citenamefont{{Williams}}(1994)}]{williams}
\bibinfo{author}{\bibfnamefont{F.~A.} \bibnamefont{{Williams}}},
  \emph{\bibinfo{title}{{Combustion Theory}}} (\bibinfo{publisher}{Westview
  Press}, \bibinfo{year}{1994}).

\bibitem[{\citenamefont{{Landau} and {Lifshitz}}(1987)}]{landau}
\bibinfo{author}{\bibfnamefont{L.~D.} \bibnamefont{{Landau}}} \bibnamefont{and}
  \bibinfo{author}{\bibfnamefont{E.~M.} \bibnamefont{{Lifshitz}}},
  \emph{\bibinfo{title}{{Fluid Mechanics}}} (\bibinfo{publisher}{London:
  Pergamon Press}, \bibinfo{year}{1987}).

\bibitem[{\citenamefont{Barton and Hodder}(1973)}]{barton1973}
\bibinfo{author}{\bibfnamefont{A.~F.~M.} \bibnamefont{Barton}}
  \bibnamefont{and} \bibinfo{author}{\bibfnamefont{A.~P.~W.}
  \bibnamefont{Hodder}}, \bibinfo{journal}{Chemical Reviews}
  \textbf{\bibinfo{volume}{73}}, \bibinfo{pages}{127} (\bibinfo{year}{1973}),
  \urlprefix\url{http://dx.doi.org/10.1021/cr60282a003}.

\bibitem[{\citenamefont{{Olinto}}(1987)}]{olinto}
\bibinfo{author}{\bibfnamefont{A.~V.} \bibnamefont{{Olinto}}},
  \bibinfo{journal}{Physics Letters B} \textbf{\bibinfo{volume}{192}},
  \bibinfo{pages}{71} (\bibinfo{year}{1987}).

\bibitem[{\citenamefont{{Heiselberg} et~al.}(1991)\citenamefont{{Heiselberg},
  {Baym}, and {Pethick}}}]{heiselberg}
\bibinfo{author}{\bibfnamefont{H.}~\bibnamefont{{Heiselberg}}},
  \bibinfo{author}{\bibfnamefont{G.}~\bibnamefont{{Baym}}}, \bibnamefont{and}
  \bibinfo{author}{\bibfnamefont{C.~J.} \bibnamefont{{Pethick}}},
  \bibinfo{journal}{Nuclear Physics B Proceedings Supplements}
  \textbf{\bibinfo{volume}{24}}, \bibinfo{pages}{144} (\bibinfo{year}{1991}).

\bibitem[{\citenamefont{{Olesen} and {Madsen}}(1991)}]{olesen}
\bibinfo{author}{\bibfnamefont{M.~L.} \bibnamefont{{Olesen}}} \bibnamefont{and}
  \bibinfo{author}{\bibfnamefont{J.}~\bibnamefont{{Madsen}}},
  \bibinfo{journal}{Nuclear Physics B Proceedings Supplements}
  \textbf{\bibinfo{volume}{24}}, \bibinfo{pages}{170} (\bibinfo{year}{1991}).

\bibitem[{\citenamefont{{Bethe} and {Johnson}}(1974)}]{bethe1974}
\bibinfo{author}{\bibfnamefont{H.~A.} \bibnamefont{{Bethe}}} \bibnamefont{and}
  \bibinfo{author}{\bibfnamefont{M.~B.} \bibnamefont{{Johnson}}},
  \bibinfo{journal}{Nuclear Physics A} \textbf{\bibinfo{volume}{230}},
  \bibinfo{pages}{1} (\bibinfo{year}{1974}).

\bibitem[{\citenamefont{{Niebergal} et~al.}(2010)\citenamefont{{Niebergal},
  {Ouyed}, and {Jaikumar}}}]{brian}
\bibinfo{author}{\bibfnamefont{B.}~\bibnamefont{{Niebergal}}},
  \bibinfo{author}{\bibfnamefont{R.}~\bibnamefont{{Ouyed}}}, \bibnamefont{and}
  \bibinfo{author}{\bibfnamefont{P.}~\bibnamefont{{Jaikumar}}},
  \bibinfo{journal}{\prc} \textbf{\bibinfo{volume}{82}},
  \bibinfo{pages}{062801} (\bibinfo{year}{2010}), \eprint{1008.4806}.

\bibitem[{\citenamefont{{Herzog} and {R{\"o}pke}}(2011)}]{herzog}
\bibinfo{author}{\bibfnamefont{M.}~\bibnamefont{{Herzog}}} \bibnamefont{and}
  \bibinfo{author}{\bibfnamefont{F.~K.} \bibnamefont{{R{\"o}pke}}},
  \bibinfo{journal}{\prd} \textbf{\bibinfo{volume}{84}},
  \bibinfo{pages}{083002} (\bibinfo{year}{2011}), \eprint{1109.0539}.

\bibitem[{\citenamefont{{Pagliara} et~al.}(2013)\citenamefont{{Pagliara},
  {Herzog}, and {R{\"o}pke}}}]{pagliara13}
\bibinfo{author}{\bibfnamefont{G.}~\bibnamefont{{Pagliara}}},
  \bibinfo{author}{\bibfnamefont{M.}~\bibnamefont{{Herzog}}}, \bibnamefont{and}
  \bibinfo{author}{\bibfnamefont{F.~K.} \bibnamefont{{R{\"o}pke}}},
  \bibinfo{journal}{\prd} \textbf{\bibinfo{volume}{87}}, \bibinfo{eid}{103007}
  (\bibinfo{year}{2013}), \eprint{1304.6884}.

\bibitem[{\citenamefont{Bhattacharyya et~al.}(2006)\citenamefont{Bhattacharyya,
  Ghosh, Joarder, Mallick, and Raha}}]{bhattacharyya}
\bibinfo{author}{\bibfnamefont{A.}~\bibnamefont{Bhattacharyya}},
  \bibinfo{author}{\bibfnamefont{S.~K.} \bibnamefont{Ghosh}},
  \bibinfo{author}{\bibfnamefont{P.~S.} \bibnamefont{Joarder}},
  \bibinfo{author}{\bibfnamefont{R.}~\bibnamefont{Mallick}}, \bibnamefont{and}
  \bibinfo{author}{\bibfnamefont{S.}~\bibnamefont{Raha}},
  \bibinfo{journal}{Phys. Rev. C} \textbf{\bibinfo{volume}{74}},
  \bibinfo{pages}{065804} (\bibinfo{year}{2006}).

\bibitem[{\citenamefont{{Mishustin} et~al.}(2014)\citenamefont{{Mishustin},
  {Mallick}, {Nandi}, and {Satarov}}}]{mishustin14}
\bibinfo{author}{\bibfnamefont{I.}~\bibnamefont{{Mishustin}}},
  \bibinfo{author}{\bibfnamefont{R.}~\bibnamefont{{Mallick}}},
  \bibinfo{author}{\bibfnamefont{R.}~\bibnamefont{{Nandi}}}, \bibnamefont{and}
  \bibinfo{author}{\bibfnamefont{L.}~\bibnamefont{{Satarov}}},
  \bibinfo{journal}{ArXiv e-prints}  (\bibinfo{year}{2014}),
  \eprint{1410.8322}.

\bibitem[{\citenamefont{{Cho} et~al.}(1994)\citenamefont{{Cho}, {Ng}, and
  {Speliotopoulos}}}]{cho}
\bibinfo{author}{\bibfnamefont{H.~T.} \bibnamefont{{Cho}}},
  \bibinfo{author}{\bibfnamefont{K.}~\bibnamefont{{Ng}}}, \bibnamefont{and}
  \bibinfo{author}{\bibfnamefont{A.~D.} \bibnamefont{{Speliotopoulos}}},
  \bibinfo{journal}{Physics Letters B} \textbf{\bibinfo{volume}{326}},
  \bibinfo{pages}{111} (\bibinfo{year}{1994}), \eprint{arXiv:astro-ph/9305006}.

\bibitem[{\citenamefont{{Tokareva} and {Nusser}}(2006)}]{tokareva}
\bibinfo{author}{\bibfnamefont{I.}~\bibnamefont{{Tokareva}}} \bibnamefont{and}
  \bibinfo{author}{\bibfnamefont{A.}~\bibnamefont{{Nusser}}},
  \bibinfo{journal}{Physics Letters B} \textbf{\bibinfo{volume}{639}},
  \bibinfo{pages}{232} (\bibinfo{year}{2006}), \eprint{arXiv:astro-ph/0502344}.

\bibitem[{\citenamefont{{Drago} et~al.}(2007)\citenamefont{{Drago}, {Lavagno},
  and {Parenti}}}]{drago}
\bibinfo{author}{\bibfnamefont{A.}~\bibnamefont{{Drago}}},
  \bibinfo{author}{\bibfnamefont{A.}~\bibnamefont{{Lavagno}}},
  \bibnamefont{and}
  \bibinfo{author}{\bibfnamefont{I.}~\bibnamefont{{Parenti}}},
  \bibinfo{journal}{\apj} \textbf{\bibinfo{volume}{659}}, \bibinfo{pages}{1519}
  (\bibinfo{year}{2007}), \eprint{arXiv:astro-ph/0512652}.

\bibitem[{\citenamefont{{Horvath}}(2010)}]{horvath2010}
\bibinfo{author}{\bibfnamefont{J.~E.} \bibnamefont{{Horvath}}},
  \bibinfo{journal}{International Journal of Modern Physics D}
  \textbf{\bibinfo{volume}{19}}, \bibinfo{pages}{523} (\bibinfo{year}{2010}),
  \eprint{arXiv:astro-ph/0703233}.

\bibitem[{\citenamefont{{Drago} and {Pagliara}}(2015)}]{drago15}
\bibinfo{author}{\bibfnamefont{A.}~\bibnamefont{{Drago}}} \bibnamefont{and}
  \bibinfo{author}{\bibfnamefont{G.}~\bibnamefont{{Pagliara}}},
  \bibinfo{journal}{ArXiv e-prints}  (\bibinfo{year}{2015}),
  \eprint{1504.02795}.

\bibitem[{\citenamefont{{Coll}}(1976)}]{coll76}
\bibinfo{author}{\bibfnamefont{B.}~\bibnamefont{{Coll}}},
  \bibinfo{journal}{Annales de L'Institut Henri Poincare Section Physique
  Theorique} \textbf{\bibinfo{volume}{25}}, \bibinfo{pages}{363}
  (\bibinfo{year}{1976}).

\bibitem[{\citenamefont{{Anile}}(1989)}]{anile89}
\bibinfo{author}{\bibfnamefont{A.~M.} \bibnamefont{{Anile}}},
  \emph{\bibinfo{title}{{Relativistic fluids and magneto-fluids : with
  applications in astrophysics and plasma physics}}} (\bibinfo{year}{1989}).

\bibitem[{\citenamefont{{Horvath} and {Benvenuto}}(1988)}]{horvath88}
\bibinfo{author}{\bibfnamefont{J.~E.} \bibnamefont{{Horvath}}}
  \bibnamefont{and} \bibinfo{author}{\bibfnamefont{O.~G.}
  \bibnamefont{{Benvenuto}}}, \bibinfo{journal}{Physics Letters B}
  \textbf{\bibinfo{volume}{213}}, \bibinfo{pages}{516} (\bibinfo{year}{1988}).

\bibitem[{\citenamefont{{Steinhardt}}(1982)}]{steinhardt}
\bibinfo{author}{\bibfnamefont{P.~J.} \bibnamefont{{Steinhardt}}},
  \bibinfo{journal}{\prd} \textbf{\bibinfo{volume}{25}}, \bibinfo{pages}{2074}
  (\bibinfo{year}{1982}).

\bibitem[{\citenamefont{Shen et~al.}(2011)\citenamefont{Shen, Toki, Oyamatsu,
  and Sumiyoshi}}]{shen11}
\bibinfo{author}{\bibfnamefont{H.}~\bibnamefont{Shen}},
  \bibinfo{author}{\bibfnamefont{H.}~\bibnamefont{Toki}},
  \bibinfo{author}{\bibfnamefont{K.}~\bibnamefont{Oyamatsu}}, \bibnamefont{and}
  \bibinfo{author}{\bibfnamefont{K.}~\bibnamefont{Sumiyoshi}},
  \bibinfo{journal}{Astrophys.J.Suppl.} \textbf{\bibinfo{volume}{197}},
  \bibinfo{pages}{20} (\bibinfo{year}{2011}), \eprint{1105.1666}.

\bibitem[{\citenamefont{Sugahara and Toki}(1994)}]{sugahara94}
\bibinfo{author}{\bibfnamefont{Y.}~\bibnamefont{Sugahara}} \bibnamefont{and}
  \bibinfo{author}{\bibfnamefont{H.}~\bibnamefont{Toki}},
  \bibinfo{journal}{Nucl.Phys.} \textbf{\bibinfo{volume}{A579}},
  \bibinfo{pages}{557} (\bibinfo{year}{1994}).

\bibitem[{\citenamefont{{Nakamura} and {Particle Data
  Group}}(2010)}]{nakamura10}
\bibinfo{author}{\bibfnamefont{K.}~\bibnamefont{{Nakamura}}} \bibnamefont{and}
  \bibinfo{author}{\bibnamefont{{Particle Data Group}}},
  \bibinfo{journal}{Journal of Physics G Nuclear Physics}
  \textbf{\bibinfo{volume}{37}}, \bibinfo{eid}{075021} (\bibinfo{year}{2010}).

\bibitem[{\citenamefont{{Farhi} and {Jaffe}}(1984)}]{farhi84}
\bibinfo{author}{\bibfnamefont{E.}~\bibnamefont{{Farhi}}} \bibnamefont{and}
  \bibinfo{author}{\bibfnamefont{R.~L.} \bibnamefont{{Jaffe}}},
  \bibinfo{journal}{\prd} \textbf{\bibinfo{volume}{30}}, \bibinfo{pages}{2379}
  (\bibinfo{year}{1984}).

\bibitem[{\citenamefont{Weissenborn et~al.}(2011)\citenamefont{Weissenborn,
  Sagert, Pagliara, Hempel, and Schaffner-Bielich}}]{weissenborn}
\bibinfo{author}{\bibfnamefont{S.}~\bibnamefont{Weissenborn}},
  \bibinfo{author}{\bibfnamefont{I.}~\bibnamefont{Sagert}},
  \bibinfo{author}{\bibfnamefont{G.}~\bibnamefont{Pagliara}},
  \bibinfo{author}{\bibfnamefont{M.}~\bibnamefont{Hempel}}, \bibnamefont{and}
  \bibinfo{author}{\bibfnamefont{J.}~\bibnamefont{Schaffner-Bielich}},
  \bibinfo{journal}{Astrophys.J.} \textbf{\bibinfo{volume}{740}},
  \bibinfo{pages}{L14} (\bibinfo{year}{2011}), \eprint{1102.2869}.

\bibitem[{\citenamefont{{Demorest} et~al.}(2010)\citenamefont{{Demorest},
  {Pennucci}, {Ransom}, {Roberts}, and {Hessels}}}]{demorest}
\bibinfo{author}{\bibfnamefont{P.~B.} \bibnamefont{{Demorest}}},
  \bibinfo{author}{\bibfnamefont{T.}~\bibnamefont{{Pennucci}}},
  \bibinfo{author}{\bibfnamefont{S.~M.} \bibnamefont{{Ransom}}},
  \bibinfo{author}{\bibfnamefont{M.~S.~E.} \bibnamefont{{Roberts}}},
  \bibnamefont{and} \bibinfo{author}{\bibfnamefont{J.~W.~T.}
  \bibnamefont{{Hessels}}}, \bibinfo{journal}{\nat}
  \textbf{\bibinfo{volume}{467}}, \bibinfo{pages}{1081} (\bibinfo{year}{2010}),
  \eprint{1010.5788}.

\bibitem[{\citenamefont{{Antoniadis} et~al.}(2013)\citenamefont{{Antoniadis},
  {Freire}, {Wex}, {Tauris}, {Lynch}, {van Kerkwijk}, {Kramer}, {Bassa},
  {Dhillon}, {Driebe} et~al.}}]{antoniadis13}
\bibinfo{author}{\bibfnamefont{J.}~\bibnamefont{{Antoniadis}}},
  \bibinfo{author}{\bibfnamefont{P.~C.~C.} \bibnamefont{{Freire}}},
  \bibinfo{author}{\bibfnamefont{N.}~\bibnamefont{{Wex}}},
  \bibinfo{author}{\bibfnamefont{T.~M.} \bibnamefont{{Tauris}}},
  \bibinfo{author}{\bibfnamefont{R.~S.} \bibnamefont{{Lynch}}},
  \bibinfo{author}{\bibfnamefont{M.~H.} \bibnamefont{{van Kerkwijk}}},
  \bibinfo{author}{\bibfnamefont{M.}~\bibnamefont{{Kramer}}},
  \bibinfo{author}{\bibfnamefont{C.}~\bibnamefont{{Bassa}}},
  \bibinfo{author}{\bibfnamefont{V.~S.} \bibnamefont{{Dhillon}}},
  \bibinfo{author}{\bibfnamefont{T.}~\bibnamefont{{Driebe}}},
  \bibnamefont{et~al.}, \bibinfo{journal}{Science}
  \textbf{\bibinfo{volume}{340}}, \bibinfo{pages}{448} (\bibinfo{year}{2013}),
  \eprint{1304.6875}.

\bibitem[{\citenamefont{{Drago} et~al.}(2014)\citenamefont{{Drago}, {Lavagno},
  and {Pagliara}}}]{drago14}
\bibinfo{author}{\bibfnamefont{A.}~\bibnamefont{{Drago}}},
  \bibinfo{author}{\bibfnamefont{A.}~\bibnamefont{{Lavagno}}},
  \bibnamefont{and}
  \bibinfo{author}{\bibfnamefont{G.}~\bibnamefont{{Pagliara}}},
  \bibinfo{journal}{\prd} \textbf{\bibinfo{volume}{89}}, \bibinfo{eid}{043014}
  (\bibinfo{year}{2014}), \eprint{1309.7263}.

\bibitem[{\citenamefont{{Lugones} and {Benvenuto}}(1998)}]{lugones98}
\bibinfo{author}{\bibfnamefont{G.}~\bibnamefont{{Lugones}}} \bibnamefont{and}
  \bibinfo{author}{\bibfnamefont{O.~G.} \bibnamefont{{Benvenuto}}},
  \bibinfo{journal}{\prd} \textbf{\bibinfo{volume}{58}}, \bibinfo{eid}{083001}
  (\bibinfo{year}{1998}).

\bibitem[{\citenamefont{{Iida} and {Sato}}(1998)}]{iida98}
\bibinfo{author}{\bibfnamefont{K.}~\bibnamefont{{Iida}}} \bibnamefont{and}
  \bibinfo{author}{\bibfnamefont{K.}~\bibnamefont{{Sato}}},
  \bibinfo{journal}{\prc} \textbf{\bibinfo{volume}{58}}, \bibinfo{pages}{2538}
  (\bibinfo{year}{1998}), \eprint{nucl-th/9808056}.

\bibitem[{\citenamefont{{Lugones} and {Horvath}}(2002)}]{lugones2002}
\bibinfo{author}{\bibfnamefont{G.}~\bibnamefont{{Lugones}}} \bibnamefont{and}
  \bibinfo{author}{\bibfnamefont{J.~E.} \bibnamefont{{Horvath}}},
  \bibinfo{journal}{\prd} \textbf{\bibinfo{volume}{66}},
  \bibinfo{pages}{074017} (\bibinfo{year}{2002}),
  \eprint{arXiv:hep-ph/0211070}.

\bibitem[{\citenamefont{{Lugones} and {Horvath}}(2003)}]{lugones2003}
\bibinfo{author}{\bibfnamefont{G.}~\bibnamefont{{Lugones}}} \bibnamefont{and}
  \bibinfo{author}{\bibfnamefont{J.~E.} \bibnamefont{{Horvath}}},
  \bibinfo{journal}{Astronomy and Astrophysics} \textbf{\bibinfo{volume}{403}},
  \bibinfo{pages}{173} (\bibinfo{year}{2003}), \eprint{arXiv:astro-ph/0211638}.

\bibitem[{\citenamefont{{Horvath} and {Lugones}}(2004)}]{horvath2004}
\bibinfo{author}{\bibfnamefont{J.~E.} \bibnamefont{{Horvath}}}
  \bibnamefont{and}
  \bibinfo{author}{\bibfnamefont{G.}~\bibnamefont{{Lugones}}},
  \bibinfo{journal}{Astronomy and Astrophysics} \textbf{\bibinfo{volume}{422}},
  \bibinfo{pages}{L1} (\bibinfo{year}{2004}), \eprint{arXiv:astro-ph/0402349}.

\bibitem[{\citenamefont{{Kov{\'a}cs} et~al.}(2009)\citenamefont{{Kov{\'a}cs},
  {Cheng}, and {Harko}}}]{kovacs2009}
\bibinfo{author}{\bibfnamefont{Z.}~\bibnamefont{{Kov{\'a}cs}}},
  \bibinfo{author}{\bibfnamefont{K.~S.} \bibnamefont{{Cheng}}},
  \bibnamefont{and} \bibinfo{author}{\bibfnamefont{T.}~\bibnamefont{{Harko}}},
  \bibinfo{journal}{Monthly Notices of the RAS} \textbf{\bibinfo{volume}{400}},
  \bibinfo{pages}{1632} (\bibinfo{year}{2009}), \eprint{0908.2672}.

\bibitem[{\citenamefont{{Lai} and {Xu}}(2009)}]{lai2009}
\bibinfo{author}{\bibfnamefont{X.~Y.} \bibnamefont{{Lai}}} \bibnamefont{and}
  \bibinfo{author}{\bibfnamefont{R.~X.} \bibnamefont{{Xu}}},
  \bibinfo{journal}{Monthly Notices of the RAS} \textbf{\bibinfo{volume}{398}},
  \bibinfo{pages}{L31} (\bibinfo{year}{2009}), \eprint{0905.2839}.

\bibitem[{\citenamefont{{Lai} and {Xu}}(2010)}]{lai2010}
\bibinfo{author}{\bibfnamefont{X.~Y.} \bibnamefont{{Lai}}} \bibnamefont{and}
  \bibinfo{author}{\bibfnamefont{R.~X.} \bibnamefont{{Xu}}},
  \bibinfo{journal}{ArXiv e-prints}  (\bibinfo{year}{2010}),
  \eprint{1011.0526}.

\end{thebibliography}

\end{document}